\documentclass[11pt,twoside]{article}
\usepackage[utf8]{inputenc}
\usepackage[T1]{fontenc}
\usepackage{amsmath, amsfonts, amssymb}
\usepackage[mathscr]{eucal}
\usepackage{mathtools}
\usepackage{times}
\usepackage{bm}
\usepackage{booktabs}
\usepackage{color}
\usepackage[natural, dvipsnames, pdftex]{xcolor}
\usepackage{algorithm}
\usepackage[noend]{algpseudocode}
\usepackage[pdftex, hidelinks]{hyperref}
\usepackage[authoryear]{natbib}
\usepackage{multirow}
\usepackage{graphicx}
\graphicspath{{figure/}}

\usepackage[top=3cm, bottom=3cm, left=2.25cm, right=2.25cm]{geometry}
\usepackage[calcwidth,  sf,  big, compact]{titlesec}
\titleformat{\section}
{\normalfont\scshape \Large}{\thesection}{1em}{}
\usepackage{fancyhdr}
\usepackage{setspace}
\usepackage[amsmath,  thref,  thmmarks,  hyperref]{ntheorem}
\usepackage{cleveref}

\titleformat{\section}[block]{\large \bf }
{  {\thesection.}}{4pt}{   }
\titleformat{\subsection}[block]{\itshape}
{  {\thesubsection.}}{4pt}{   }

\def\and{\textsc{and }}

\def\Pr{{\rm Pr}}

\def\calC{{\mathcal C}}

\def\upsilon{{\Upsilon}}
\def\bi{\begin{itemize}}
\def\ei{\end{itemize}}
\def\F{{S}}

\def\calH{{H}}
\newcommand{\D}[1]{\mathrm{d}{#1}}
\def\var{{\rm var}}

\newcommand{\Istat}{\textsf{ISTAT}}
\newcommand{\IDLold}{\textsf{IDL 2016}}

\newcommand{\IDL}{\textsf{IDL}}
\newcommand{\Rlang}{\ensuremath{\mathsf{R}}{}}

\title{Is there a cap on longevity? A statistical review}

\author{L\'eo R. Belzile\thanks{HEC Montréal, Department of Decision Sciences, 3000, chemin de la Côte-Sainte-Catherine, Montréal (QC), Canada H3T 2A7 (\texttt{leo.belzile@hec.ca})}, Anthony C. Davison\thanks{EPFL-SB-MATH-STAT, Station 8, \'Ecole polytechnique f\'ed\'erale de Lausanne, 1015 Lausanne, Switzerland}, Jutta Gampe\thanks{Laboratory of Statistical Demography, Max Planck Institute for Demographic Research, Rostock, 18057, Germany}, Holger Rootzén and  Dmitrii Zholud\thanks{Department of Mathematical Sciences, Chalmers and Gothenburg University, Chalmers Tvärgata 3, 41296 Göteborg, Sweden}
}
\date{}

\begin{document}

\maketitle

%Abstract
\begin{abstract}
There is sustained and widespread interest in understanding the limit, if any, to the human lifespan.  Apart from its intrinsic and biological interest, changes in survival in old age have implications for the sustainability of social security systems. A central question is whether the endpoint of the underlying lifetime distribution is finite.  Recent analyses of data on the oldest human lifetimes have led to competing claims about survival and to some controversy, due in part to incorrect statistical analysis.  This paper discusses the particularities of such data, outlines correct ways of handling them and presents suitable models and methods for their analysis.  We provide a critical assessment of some earlier work and illustrate the ideas through reanalysis of semi-supercentenarian lifetime data.  Our analysis suggests that remaining life-length after age 109 is exponentially distributed, and that any upper limit lies well beyond the highest lifetime yet reliably recorded.  Lower limits to 95\% confidence intervals for the human lifespan are around 130 years, and point estimates typically indicate no upper limit at all.
\end{abstract}
% 
% %Keywords, etc.
% \begin{keywords}
% Censoring, data validation, generalized Pareto distribution, Gompertz distribution, hazard function, Lexis diagram, survival analysis, supercentenarian, truncation
% \end{keywords}

\section{INTRODUCTION}

The possible existence of a hard upper limit, a cap, on human lifetimes is hotly debated and attracts widespread public interest. Answers to the question are important demographically  and for pension systems, since the presence of a {cap} may imply that the upward trend in expected lifetimes that has occurred over the last century in developed countries cannot continue. Moreover, establishing the existence or not of such a limit would inform biological theories of ageing and could aid efforts to prolong lives.

In this paper we discuss statistical approaches to understanding extreme human lifetimes and to investigating whether a limit exists.  Several  components to this interact: the availability and quality of data, statistical models and methods for addressing the issue appropriately, and extrapolation. We illustrate these issues by reanalysing data on extreme human lifetimes, give a critical review of earlier work, and present our view on the debate.  Throughout we use the term `(human) lifespan' for the generic, theoretical, limit to the lives of humans, and the term `lifetime' to refer to the life-length of an individual human. 

The first component, data sources and data quality, is discussed in Section~\ref{sec:data}. In statistical terms our opening question --- \textsl{Is there a cap on longevity?} --- corresponds to investigating whether the support of the lifetime distribution is bounded or unbounded, and, more practically, how this may be inferred from data. One key aspect, crucial not just in the present context but definitely a central issue here, is whether the available data are representative of the population under study. This is the first aspect discussed in Section~\ref{sec:data}.  As the research question requires information on the right-hand tail of the lifetime distribution, reliable analyses depend critically on the quality of data on unusually long lifetimes.  Extreme ages at death are well above 100 years and, although many developed countries have long high-quality records of births and deaths, this implies that we must rely on data that go back more than a century. Continuous archiving may have been obstructed by wars, changes of territory or political systems, and even complete data series may be tarnished by errors of entry and transcription that compromise studies of very old individuals.  Data validation is also addressed in Section~\ref{sec:data}. In addition to representative sampling and data validation, a further complication is the effect of different observation schemes.

Section 3 sketches methods for the analysis of lifetime data. A central function in survival analysis is the hazard of death, or `force of mortality' as it is called in demography. This is the failure rate of the lifetime distribution; it quantifies the risk of death in the next infinitesimal interval, conditional on survival thus far. For random variables modeling lifetimes with finite support the hazard necessarily increases to infinity when lifetimes approach the upper boundary, which is then sometimes very expressively called a `{wall of death}'. A hazard that increases to infinity, however, need not impose a finite lifespan, and this has led to some confusion.  A lifetime distribution whose hazard function is bounded must have an infinite upper limit. 

Rounding, censoring and truncation are very common with lifetime data; the first two reduce information on the observations, and the third determines which observations are available at all.  In  Section 2 we illustrate the effect of truncation and in Section 3 we discuss methods for removing the resulting biases. 

Section 3 also reviews the analysis of extreme lifetimes, including the use of the generalized Pareto distribution arising in extreme value statistics and  the Gompertz distribution commonly used in demography,  and discusses how heavy rounding, truncation, and  censoring impact estimation of the hazard.

Most statistical analyses involve some form of extrapolation, from a sample to a population or from the past to the future.   Statistical  extrapolation entails giving some idea of the underlying uncertainty, formal statement of which requires  a stochastic model.  Semi- and non-parametric models are complex, flexible and widely used in lifetime data analysis but are unreliable outside the range of the data, whereas parametric models are simpler and allow extrapolation, though this may be biased if the model fits the data poorly.  Here our focus is extrapolation beyond the data to a possible limiting lifespan, and this requires parametric modeling, but the fit of competing models must be carefully assessed, alongside the range of ages over which they are adequate.  

There is no doubt that the force of mortality increases up to at least 105 years, so if one uses a parametric model to analyze datasets dominated by lifetimes around 100 years, the fitted model will reflect this increase. For the generalized Pareto model this necessarily leads to a finite lifespan estimate, whereas for the Gompertz model, which is widely fitted at lower ages,  the force of mortality increases indefinitely but the estimated lifespan is infinite. Such extrapolations were the only ones possible before the existence of the International Database on Longevity (\IDL{}), but they agree badly with the \IDL{} supercentenarians, who show a plateauing of the force of mortality.  Hence extrapolations from ages around 100 to ages 120 or higher  cannot now be regarded as trustworthy.  Caveats are needed for extrapolations far beyond the \IDL{} data range of 105--122 years.

In Section 4 we illustrate the consequences of extrapolation from ages around 100 and of rounding and truncation. `Plateauing'  is a central issue at higher ages: does the force of mortality becomes constant at some point, or does it continue to increase, perhaps so much that a finite lifespan is effectively imposed?   We review the part of the literature which concludes that there in fact is plateauing, perhaps around age 109, and note that this agrees with our analyses. We also review differences in mortality at extreme age between gender, between persons born earlier or later, and between  countries, adding results for newly-available \IDL{} data. The  impression is that such differences are limited.

In Section 5 we discuss the analysis strategies in some of the papers which find a cap on the human lifespan. Key issues here are the handling of truncation and censoring and extrapolation from relatively low to much higher ages. Our overall conclusions are given in Section 6.

\section{DATA ON EXTREME LIFETIMES}\label{sec:data}

\subsection{General}

Data on extreme ages at death are needed to investigate the upper tail of the lifetime distribution. Such data must be collected to reflect the actual tail of the lifetime distribution and not a distorted version of it. In statistics this is termed representative sampling, and it requires a well-defined and correctly-implemented observation scheme.  Observations  on individuals still alive at very high ages could also be used if they stem from a well-defined sample, but access to such data is rare.

A key element is the age range that is deemed to contain the extreme lifetimes, that is, which ages form the upper tail. Supercentenarians, who survive to or beyond their 110$^\text{th}$ birthdays, are commonly considered to constitute this group. If we extend the age range downward by five years, then we may include semi-supercentenarians, who live to see their 105$^\text{th}$ but not their 110$^\text{th}$ birthdays.

Information on potential (semi-)supercentenarians should be treated with caution. Age-overstatement is all too frequent, as a very long life is highly-respected, so data on supercentenarians must be carefully and individually validated to verify that the reported age at death is correct. Such validation ideally requires a check of a birth or baptism records, preferably supplemented by early-life documents such as early census records or marriage certificates, that can be linked conclusively to the individual. The documents to be used depend on the national administrative system.  Not surprisingly, age validation of foreign-born individuals is often difficult,  so studies are frequently limited to native-born persons. \cite{Poulain:2010} discusses age-validation in detail.

\subsection{Data resources}

There are two main resources for validated supercentenarian lifetimes: the International Database of Longevity (\IDL{}) and the collection of the  Gerontology Research Group. 
They can be found online at \texttt{www.supercentenarians.org} and at \texttt{www.grg.org}.  As they are updated often, care is needed when attempting to reproduce previous analyses.
%\bluefn{changing}

The \IDL{} contains data assembled by government agencies. The data collection varies by country but is always based on a well-defined and documented procedure that should be respected in any statistical analysis. This is to avoid age-ascertainment bias, which arises if the inclusion of an individual in the sample depends, in an unspecified way, on his or her age. Such bias is typically present if record-holders or individuals with extreme ages are more likely to appear in a dataset than younger ones. As this unequal sampling is not quantifiable, the resulting distortion cannot be managed.

The \IDL{} was originally compiled to allow the estimation of the trajectory of human mortality, i.e., the hazard of death, up to the highest ages \citep{IDL:2010}. Initially the \IDL{} provided  data only on supercentenarians, but it now  includes semi-supercentenarians for some countries \citep{IDL:2021}. At the time of writing it contains  validated supercentenarian lifetimes from 13 countries, and information on semi-supercentenarians for nine countries. 

The Gerontology Research Group website provides lists of known and validated supercentenarians, including some currently alive: on 1 April 2021, for example, the database stated that the oldest living validated person, Kane Tanaka, was aged 118 years and 94 days. It contains many interesting facts, but age-ascertainment bias and the absence of a specified observation scheme render it unsuitable for statistical analysis.

Certain other databases that qualify as ascertainment-bias-free samples of long-lived individuals and have been used in lifetime analyses are discussed in {Section~4}.

\subsection{Observation schemes}

Creating a sample usually starts by `casting a net' \citep{Kestenbaum.Ferguson:2010} on the population of potential (semi)-supercentenarians. The candidates identified in this step are then individually validated.  The population on which the net is cast for the most part comprises deceased individuals who died above a certain age $u_0$, here 110 or 105 years. The specific observation scheme depends on how the limits of the net are defined.

\begin{figure}[t!] 
 \centering 
 \includegraphics[width=\textwidth]{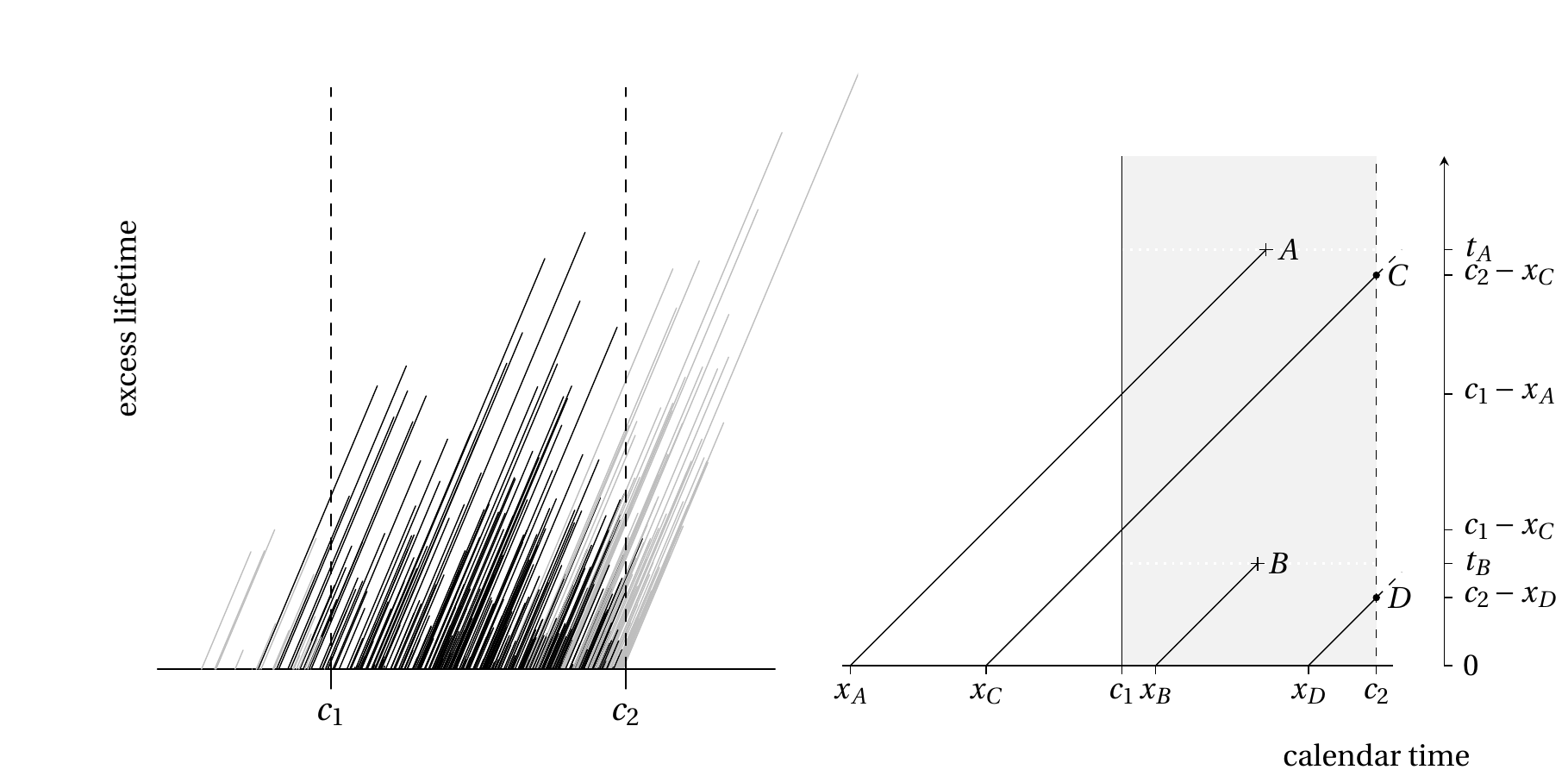}
 \caption{Truncation and the Lexis diagram.  Left: life trajectories sampled under interval truncation; only individuals with black trajectories appear in the sample. Right: schematic showing life trajectories of individuals subject to left truncation and possible right censoring; those who would attain age $u_0$ after calendar time $c_2$ or those who died before calendar time $c_1$ or below age $u_0$ are unobserved. Individual $A$ passes age $u_0$ at date $x_A<c_1$, has excess lifetime (beyond $u_0$) of $c_1-x_A$ at calendar time $c_1$,  and dies at age $u_0+t_A$; she is left-truncated at $c_1-x_A$. Individual $B$ reaches age $u_0$ at calendar time $x_B$ and dies between calendar times $c_1$ and $c_2$ at age $u_0+t_B$.  Individual $C$ is left-truncated at excess lifetime $c_1-x_C$ and right-censored at excess lifetime $c_2-x_C$. Individual $D$ is right-censored at excess lifetime $c_2-x_D$.  If the data were right-truncated, then individuals $C$ and $D$, who do not die in the shaded region $\calC=[c_1,c_2]\times[u_0,\infty)$, would not appear in the sample.}
 \label{Lexis.fig}
 \end{figure}

One popular design is the following: all individuals who died at or above age $u_0$ between two calendar dates $c_1$ and $c_2$ are collected and validated.  Individual excess lifetimes above $u_0$ are depicted in the Lexis diagram \citep{Keiding:1990}, in which each individual lifetime is shown as a diagonal line of unit slope, commencing at a calendar time $x$ on the horizontal axis and later ending in death or censoring at calendar time $x+t$.  The left-hand panel of Figure~\ref{Lexis.fig} illustrates the effect of this observation scheme for a hypothetical population in which lifetimes are sampled from a fixed distribution  but the number of individuals living beyond age $u_0$ grows over time, as in many countries.  

\begin{figure}[!t]
%\begin{figure}[t!]  
  \centering
  \includegraphics[width=0.925\textwidth]{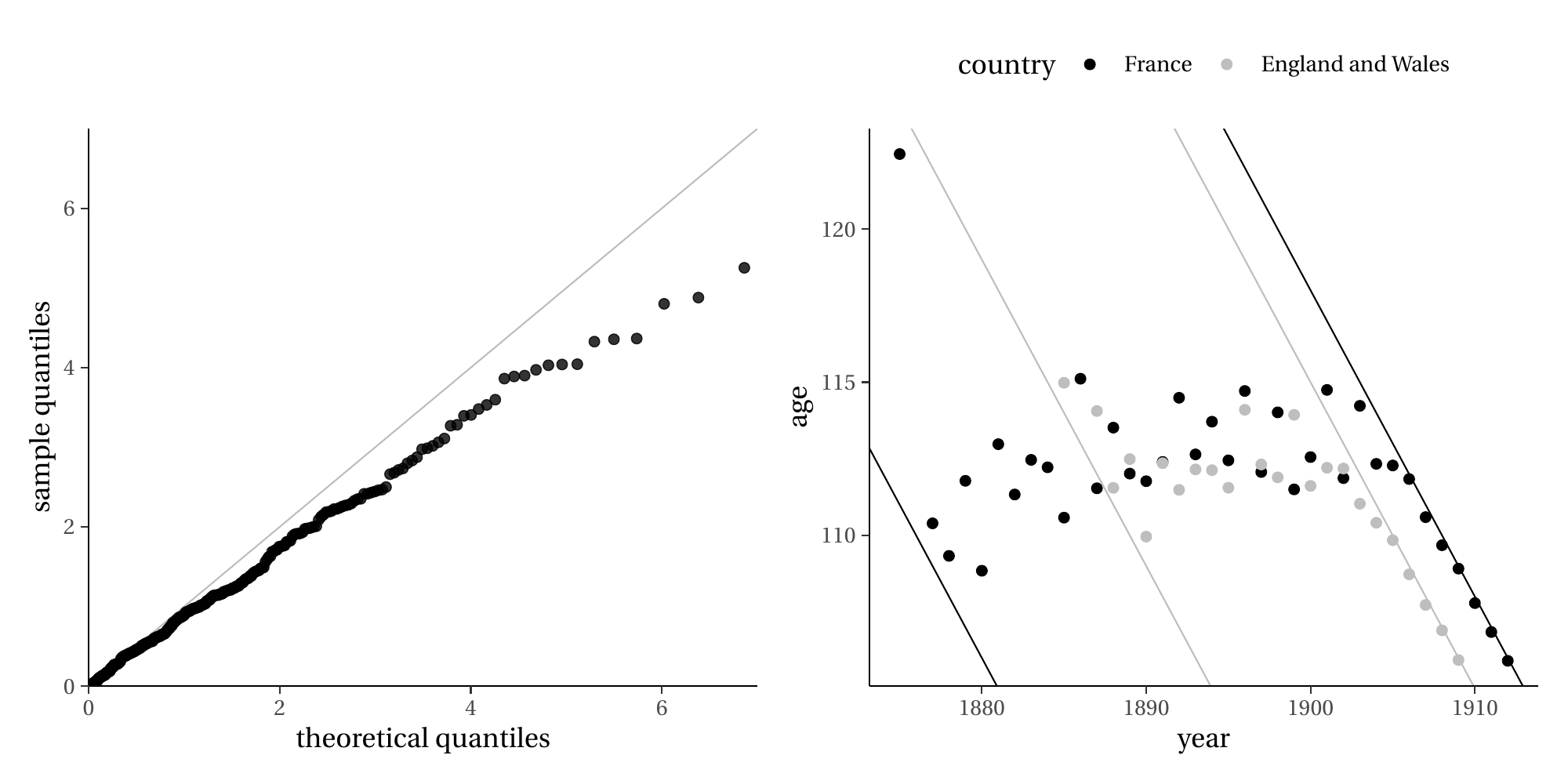}\\
  \caption{Effects of truncation on sampled lifetimes. Left: quantile-quantile plot of the truncated observations shown in the left-hand panel of \Cref{Lexis.fig} against plotting positions for the known lifetime distribution. Right: maximum reported age at death of \IDL{} individuals for France and England \&~Wales  by birth cohort. The sloping lines show the minimum and maximum possible reported ages given the sampling intervals $[c_1,c_2]$.   As the individuals must be aged at least 105, the maximum age at death for the French birth cohort 1910, say, is just under 108, since for these data $c_2$ corresponds to 31 December 2017.}
  \label{fig:truncationeffects}
\end{figure}

Conclusions will be biased if the analysis does not take the observation scheme into account. For the truncated data depicted in \Cref{Lexis.fig}, for death occurring close to $c_1$, only individuals who survived beyond the age they attained at time $c_1$ are included, so longer lifetimes are over-represented. Close to $c_2$ only individuals who died before time $c_2$ appear in the sample. Although the sample does not represent the distribution of excess lifetimes per se, the distortion due to truncation can be removed because the observation scheme is known; see Section~\ref{sec:censtrunc}. 

The left-hand panel of \Cref{fig:truncationeffects} compares the observed excess lifetimes with the true lifetime distribution. It suggests that, other things being equal, the shortening of the sample tails increases if there is an increasing trend in the numbers of individuals reaching $u_0$. This is also apparent when looking at maximum age achieved by birth cohort, where the effect of truncation leads to a downward trend, as illustrated in the right-hand panel of \Cref{fig:truncationeffects}. As recent \IDL{} releases include both supercentenarians and semi-supercentenarians, other truncation patterns arise that are discussed in  \Cref{sec:appendixA}.

\section{STATISTICAL MODELING}

Statistical analysis of extreme lifetimes rests on survival analysis and extreme value models.  The first is discussed in Sections~\ref{sec:SA}--\ref{sec:Humanlife}. We introduce basic concepts and discuss the complications (censoring and truncation) that certain observation schemes entail and how they can be adjusted for by correctly setting up the likelihood. Although the emphasis here is on parametric models we also discuss nonparametric estimation (Section~\ref{sec:npestim}) and how censored and truncated data can be handled in this case. We illustrate specific problems of observation schemes that have gone unnoticed in previous research on lifespan limits in Section~\ref{sec:samplingscheme} and give an overview of the most popular models for (old-age) human mortality in Section~\ref{sec:Humanlife}. Section~\ref{sec:EVM} presents threshold exceedance models from extreme value theory. The pivotal role of the generalized Pareto distribution is highlighted (Section~\ref{GPD.sect}), and penultimate approximations and their importance in the assessment of tail behavior are discussed in Section~\ref{sec:penultimate}. Extended models are presented in \Cref{sec:extGPD}.

\subsection{Elements of survival analysis}\label{sec:SA}
\subsubsection{Basic notions}  \label{sect.survival-basics}
Survival analysis began with John Graunt's \citeyear{Graunt:1662}  actuarial life table and remains in full development.
Systematic accounts may be found in books such as \citet{Cox.Oakes:1984}, \citet{Andersen.Borgan.Gill.Keiding:1993}, \citet{Therneau.Grambsch:2000}, \citet{Aalen.Borgan.Gjessing:2008} and \citet{Moore:2016}.

Consider a positive continuous random variable $T$ with distribution and density functions $F$ and $f$, and let $t_F=\sup\{t: F(t)<1\}$ denote the largest possible value of $T$, which may be finite or infinite.  The survivor and hazard functions $\F(t)=1-F(t)$ and $h(t)=f(t)/\F(t)$ play central roles in survival analysis and satisfy
\begin{align*}
-\log\F(t) = \int_0^t h(x)\,\D{x} = \calH(t), 
\end{align*}
where the cumulative hazard function $\calH$ should tend to infinity as $t\to t_F$; if not, $\F(t)$ does not tend to zero, implying that the event $T=t_F$ has a positive probability, sometimes called the cure fraction if $t_F=\infty$, in which case it corresponds to long-term survival.  If $t_F$ is finite and the cure fraction is zero, then $h(t)$ must increase sufficiently rapidly that $\calH(t)\to\infty$ as $t\to t_F$. 

The hazard function $h(t)$, known as the force of mortality in demography, is the rate of failure in the infinitesimal interval $[t,t+\D{t})$ conditional on survival to $t$.  It plays a central role in survival analysis.  The simplest possibility is that $h(t)$ equals a positive constant $\lambda$, and then $T$ is exponentially distributed with mean $\lambda^{-1}$.  Other possibilities are decreasing or increasing failure rate, whereby $h(t)$ is monotone decreasing or monotone increasing in $t$, or a convex `bathtub' shape. 

Discrete survival data can arise when continuous responses are heavily rounded, and in nonparametric estimation of quantities defined for continuous data; see \Cref{sec:npestim}.  
In discrete time failures are supposed to be possible at times $0\leq t_1<t_2<\cdots$ with positive probabilities $f_j=\Pr(T=t_j)$. In this case we define the hazard function by
\begin{align}\label{discrete.eq1}
h_j = \dfrac{f_j}{f_j+f_{j+1}+\cdots}, \quad\Pr(T>t_j\mid T\geq t_j) = 1-h_j,
\end{align}
can write
\begin{align}\label{discrete.eq2}
f_j = h_j\prod_{i=1}^{j-1} (1-h_i),  \quad 
\F(t) = \prod_{i:t_i<t} (1-h_i), \quad t_j<t\leq t_{j+1},
\end{align}
and define the cumulative hazard function as $\calH(t) = -\sum_{i:t_i<t} \log(1-h_i)$ for compatibility with the continuous case; when the $h_i$ are small one may replace $-\log(1-h_i)$ by $h_i$, giving the more natural definition $\calH(t) = \sum_{i:t_i<t} h_i$.

Maximum likelihood estimation and accompanying inferential tools such as likelihood ratio testing are standard in survival analysis.  They provide a well-understood inferential framework that is readily adapted to allow for censoring and truncation, and can be regarded as approximations to Bayesian inferences with vague priors, which may be preferred by some investigators.  We illustrate both likelihood and Bayesian methods below.

\subsubsection{Censoring and truncation}\label{sec:censtrunc}

Survival times are often subject to partial observation.  The simplest form of this is right censoring, whereby $T$ is supplemented with an indicator variable $\delta$, with $\delta=1$ if $T$ is observed to take value $t$, and $\delta=0$ if $T$ is right-censored, i.e., it is known only that $T>t$.  In this case the likelihood contribution can be written as $f(t)^\delta\F(t)^{1-\delta}=h(t)^\delta\F(t) $.   Left censoring and interval censoring can also arise; a left-censored response is known only to satisfy $T<t$, whereas an interval-censored response is known to fall into an interval $\mathcal{I}$, and would give a likelihood contribution $\Pr(T\in\mathcal{I})$. In the present setting, interval censoring could arise when the age at death in years, $a$, but not in days, is provided, so an individual has lifetime $T$ falling into $\mathcal{I}=[a,a+1)$ years. 

Truncation determines which observations are available for analysis, whereas censoring reduces the information that can be extracted from them.   Under truncation, a unit appears in the dataset only if its response variable falls into a subset, often an interval $\mathcal{I}$, of its possible values.  In this case the likelihood contribution is of the form $f(t\mid T\in\mathcal{I})$, with the conditioning expressing the restriction that $T$ is available only if it lies in  $\mathcal{I}$.  

In the present context lifetimes are commonly interval-truncated, i.e., truncated to both left and right, or left-truncated and right-censored.  The first arises when individuals are only included in a sample if they die within an interval $\mathcal{I}=[a,b]$, say, and the second arises when individuals who do not die within $\mathcal{I}$ are known to be alive at its end.  The corresponding likelihood contributions may respectively be written as
\begin{align}\label{lik-contribs.eq}
\dfrac{f(t)}{\F(a)-\F(b)},\quad a<t<b,\qquad \dfrac{h(t)^\delta\F(t)}{\F(a)}, \quad t>a.
\end{align}
where the possible observations $(t,\delta)$ in the second case are $(b,0)$ if $t$ is right-censored at $b$ and $(t,1)$ if $t$ is fully observed; then $a<t<b$. See also~\Cref{sec:samplingscheme}.   Care is needed when analysing the \IDL{} lifetimes, since the truncation bounds for the semi-supercentenarians and supercentenarians may differ; this is discussed in \Cref{sec:appendixA}. 

\subsubsection{Nonparametric estimation}\label{sec:npestim}

Nonparametric estimates of the survival and cumulative hazard functions cannot be used to extrapolate outside the observed data and so cannot directly address the issue of a cap on longevity, but can be used to compare the fits of competing parametric models, for example through quantile-quantile plots that account for the observation scheme \citep{Waller/Turnbull:1992}.  Nonparametric estimates can only place mass on the distinct observed failure times, as the data themselves contain no evidence that failure is possible at any other time. 

Consider a random sample of survival times $t_1, \ldots,  t_n$ from a distribution $F$, where it is supposed that $a_i<t_i<b_i$ and the truncation limits $a_i$ and $b_i$ equal zero and infinity if $t_i$ is not truncated. With no censoring or truncation, the nonparametric maximum likelihood estimate $\hat{F}$ of $F$ places masses $1/n$ on each of the $t_i$, and the survivor and cumulative hazard functions can be estimated through~\Cref{discrete.eq1,discrete.eq2}. Modification of $\hat{F}$ to allow for left-truncation and right-censoring yields the Kaplan--Meier or product-limit estimator  \citep{Kaplan.Meier:1958,Tsai/Jewell/Wang:1987}, which is most simply expressed in terms of the number of failures $d_j\in \{0, \ldots, r_j\}$ at the distinct failure times $t'_1<\cdots< t'_J$ and the number $r_j$ of individuals still at risk, i.e.,  all individuals not yet failed or censored at $t'_j$.  Then $\hat{h}_j = d_j/r_j$ and 
\begin{align*}
\hat\F(t) = \prod_{j:t_j'<t} (1-\hat{h}_j),
\quad \var\left\{\hat\F(t) \right\} \doteq  \hat\F(t)^2 \sum_{j:t'_j<t} \dfrac{\hat{h}_j}{r_j(1-\hat{h}_j)}, \quad t>0.
\end{align*}
\citet[][Section~IV]{Andersen.Borgan.Gill.Keiding:1993} review methods for obtaining pointwise and simultaneous confidence intervals for nonparametric estimators of the hazard and survivor functions.

There are no general explicit formulae for estimators under more complex observation schemes.  One approach for right-censored and interval-truncated data applies the EM algorithm  \citep{Dempster.Laird.Rubin:1977}. Let the random variable $T_i$ taking value $t_i$ have truncation set $\mathcal{T}_i\subset\{1, \ldots,  J\}$; this is the set of $j$ for which $T_i=t_j'$ is possible. Likewise define a censoring set $\calC_i\subset\mathcal{T}_i$ containing all the indices $j$ for which $T_i=t_j'$ is possible; if $T_i$ is observed  then $\calC_i$ contains a single element, but if $T_i$ is censored then $\calC_i$ has the form $\{j_1, \ldots,  j_2\}$.    We associate indicator variables $\gamma_{ij}$ and $\tau_{ij}$ to the censoring and truncation sets: $\gamma_{ij}=1$ if and only if $j \in \calC_i$ and $\tau_{ij}=1$ if and only if $j \in \mathcal{T}_i$, for $i=1, \ldots, n$ and $j = 1, \ldots ,J$.  If the $T_i$ can only take values in  $t'_1, \ldots, t'_J$, then nonparametric maximum likelihood estimation of $F$ involves finding the probabilities $f_j = \Pr(T=t'_j) \geq 0$ that maximise the likelihood 
\begin{align}
\prod_{i=1}^n \dfrac{\sum_{j=1}^J \gamma_{ij}f_j}{\sum_{j=1}^J \tau_{ij} f_j} \label{eq:emnpsurv}
\end{align}
 subject to $\sum_{j=1}^J f_j=1$ \citep{Turnbull:1976}.  The likelihood might be maximised directly if $J$ is very small, but it is typically preferable to note that if the $T_i$  had been observed without censoring or truncation then the log likelihood written in terms of indicator variables $I_{ij} = I(t_i=t_j')$ would be $\sum_{i=1}^n \sum_{j=1}^J I_{ij}\log f_j$, 
yielding $\hat{f}_j=\sum_i I_{ij}/\sum_{i,j} I_{ij}$;
this is the M~step of an EM algorithm.  The E~step involves replacing $I_{ij}$ at the $m$th iteration by
\begin{align}\label{EM.eq}
\hat{I}_{ij}^m =  \dfrac{\gamma_{ij}\hat{f}_j^m}{\sum_{j'=1}^J \gamma_{ij'} \hat{f}_{j'}^m} +  \dfrac{(1-\tau_{ij})\hat{f}_j^m}{ \sum_{j'=1}^J \tau_{ij'} \hat{f}_{j'}^m} ,
\quad i=1, \ldots,  n,\,  j=1, \ldots,  J, 
\end{align}
where $\hat{f}_1^m, \ldots,  \hat{f}_J^m$ result from the M~step. The two terms in \Cref{EM.eq} correspond to the probability that a possibly censored $T_i$ equals $t_j'$ and to `ghosts', i.e., individuals who would have been observed had there been no truncation.   The algorithm consists of setting $\hat{f}_1^0=\ldots=\hat{f}_J^0=1/J$, computing the $\hat{I}_{ij}^1$ by the E~step and then $\hat{f}^1_1, \ldots,  \hat{f}^1_J$ by the M~step, increasing $m$, and then alternating the E~and M~steps until convergence.  The observed information is found by differentiating the logarithm of \Cref{eq:emnpsurv}. 

\subsubsection{Diagnostic plots} \label{subsec:diagplot}

Standard graphical diagnostics must be modified to account for different observation schemes. If $F_0$ is an estimated or postulated parametric distribution and $F_n$ is a nonparametric estimator of the distribution function, we can construct a Q-Q plot by plotting observed failure times $t_i$ against the positions $v_i = F_0^{-1}\{F_n(t_i)\}$ on the $x$-axis. With truncated data, $a_i < t_i <b_i$, so each observation has a different distribution, yielding $v_i=F_0^{-1}[F_0(a_i) +\{F_0(b_i)-F_0(a_i)\}\{F_n(t_i) - F_n(a_i)\}/\{F_n(b_i)-F_n(a_i)\}]$.  One effect of truncation is that  the ranks of $v_i$ and $t_i$ need not coincide, so the graph need not be monotone.  Approximate confidence intervals  may be obtained by parametric bootstrapping \citep{Belzile:2020}; see the {sec:appendixD}.

\subsection{Observation schemes and likelihoods} \label{sec:samplingscheme}

We now discuss how likelihood contributions depend on the underlying observation scheme.  Consider the right-hand panel of Figure~\ref{Lexis.fig}, in which the trajectory of each individual lifetime is shown as a diagonal line of unit slope.  Let $\calC$ denote the region $[c_1,c_2]\times[u_0,\infty)$, where the calendar times $c_1$ and $c_2$ are the sampling limits.  Individuals whose trajectories do not intersect $\calC$ are unobserved; indeed, their existence cannot be inferred from the available data.   

Let $\mathcal{T}=\{(x+s,u_0+s):s>0\}$ be the trajectory of someone who passes age $u_0$ at calendar time $x$, and let $T$ denote her excess lifetime after $u_0$.  For example, if $u_0$ corresponds to 105 years then $x$ would be the calendar date of her 105th birthday, and if her excess lifetime took value $T=t$, then she would trace the line from $(x,u_0)$ to $(x+t,u_0+t)$.    

If there is interval truncation, then the argument in \Cref{sec:censtrunc} implies that the appropriate likelihood contribution for a person with trajectory $\mathcal{T}$ who dies inside $\calC$ is $f(t\mid T\in\mathcal{I})$, where $\mathcal{I}$ denotes the projection of $\calC\cap\mathcal{T}$ onto the vertical axis in the figure, and it is then easy to check that one should take
\begin{align}\label{inteval.eq}
\mathcal{I} = [a,b] = [\max(0,c_1-x),c_2-x]
\end{align}
in the first expression in \Cref{lik-contribs.eq}.  If there is left truncation and right censoring,  then these values of $a$ and $b$ are inserted into the second expression of \Cref{lik-contribs.eq}.  Hence each individual independently contributes a conditional density term to the overall likelihood.

Truncation may be hidden.  Suppose, for example, that data for birth cohorts for years $b_{\rm min}, \ldots, b_{\rm max}$ are available but only those for years $b_{\rm min}, \ldots, b^*$ are retained, because some members of birth cohort $b^*+1$ are still alive at time $c_2$.  As only extinct cohorts are used, truncation and censoring appear to be absent. This would simplify analysis if it were true, but truncation is present nonetheless.   The boundary $c_1$ in the Lexis diagram is effectively $-\infty$, but $b^*$ has been chosen using the survival times, as the corresponding random variable $B$ is the last birth cohort in which no-one dies after $c_2$.  Suppose the persons in  cohorts $b_{\rm min}, \ldots, b^*$ enter  $\calC$ at dates  $x_1, \ldots, x_{n}$ and have lifetimes $T_1, \ldots,  T_{n}$ above $u_0$.  Then 
\begin{align*}
\Pr(B=b^*) = \prod_{j=1}^{n} \Pr(T_j\leq c_2-x_j)\times p,
\end{align*}
where $p$ is  the probability that at least one individual in cohort $b^*+1$ is alive at $c_2$.  The joint probability element for the observed lifetimes $t_1, \ldots,  t_n$ is $\prod_{j=1}^n f(t_j)\times p$, and conditioning on the selection event $B=b^*$ yields
\begin{align*}
f(t_1, \ldots,  t_n\mid B=b^*) = \prod_{j=1}^{n} \dfrac{f(t_j)}{F( c_2-x_j)}, \quad 0<t_j <c_2-x_j, \quad j=1, \ldots,  n,                                                                                                                    \end{align*}
which is a product of terms for truncated observations in~\Cref{lik-contribs.eq}, with $a_j=0$ and $b_j=c_2-x_j$.  Thus this approach entails truncation, though ignoring it may be harmless if $F(c_2-x_j)\approx 1$ for all $j$.  As using only extinct cohorts does not greatly simplify analysis, and the resulting smaller sample leads both to more variable estimates and to reduced power for model comparison and assessment, it cannot be recommended.  \Cref{sec:appendixB} contains a small simulation study to illustrate this. 

It is superficially appealing to analyze the excess lifetimes for those persons dying in each year, ignoring the truncation, but doing so conflates the lifetime distribution $f(t)$ and the rate $\nu(x)$ at which individuals reach age $u_0$.  The excess lifetime for an individual dying in the interval $[c_1,c_2]$ has density $f_\calC(t)$ proportional to $f(t)w(t)$, where the weighting function 
\begin{align*}
w(t) = \int_{c_1-t}^{c_2-t} \nu(x)\,\D{x} , \quad t>0, 
\end{align*}
is decreasing because medical and social advances have increased $\nu$ over a long period. Hence naive analysis of the excess lifetimes of the persons dying in $\calC$ will yield an estimate of $f_\calC(t)$, which is a tilt of $f(t)$ towards the origin; see the left-hand panel of \Cref{fig:truncationeffects}. 

In addition to taking the observation scheme into account, the use of conditional likelihood contributions such as those in~\Cref{lik-contribs.eq} has the important advantage that $\nu$ does not influence the analysis.  See~\citet{Keiding:1990} or~\citet{Davison:2018} for more discussion. 

\subsection{Models for human lifetimes}\label{sec:Humanlife}

The trajectory of age-specific mortality, i.e., the hazard of human lifetime, has proved to be remarkably stable in its overall shape, but has dropped markedly over time \citep{Burger:2012}. Parametric modelling of human mortality over the full age range is difficult due to its complex shape, but when the focus is on so-called senescent mortality, starting from mid-life, an exponential increase in human death rates was noticed by \citet{Gompertz:1825}, leading to the hazard
\begin{equation} \label{eq:Gompertzhazard}
  h(t) = \sigma^{-1} \, e^{\beta t/\sigma } , \quad t>0,\quad \beta,\sigma>0,                                             
\end{equation}
with survivor function 
\begin{align*}
\F(t) = \exp\left\{-(e^{\beta t/\sigma }-1)/\beta\right\}.   
\end{align*}
%\Cref{eq:Gompertzhazard} 
This model is typically used to describe the increase of death risks with age for ages $t$ over 50 years. \cite{Makeham:1860} added an age-independent component $\lambda > 0$ to give what is now called the Gompertz--Makeham model, 
\begin{equation} \label{GM-haz.eq}
  h(t) = \lambda +  \sigma^{-1} \, e^{\beta t/\sigma }, \qquad 
\calH(t)= \lambda t  + (e^{\beta t/\sigma }-1)/\beta , 
\end{equation}
which often proves better at lower ages. The dimensionless parameter $\beta$ determines how fast the age-specific component increases; if $\beta=0$ then this risk is constant and only the sum of the hazards, $\lambda + 1/\sigma$,  is identifiable.   Although its hazard function increases indefinitely for $\beta > 0$, the Gompertz--Makeham model imposes no finite upper limit on lifetimes.  Taking $\beta<0$ in \Cref{eq:Gompertzhazard}  yields a defective distribution, as then $\F(t)\to\exp(1/\beta)$ as $t\to\infty$.

\cite{Perks:1932} noted that \textsl{`\ldots Makeham's curves ran much too high at the older ages'}, a phenomenon now known as late-life mortality deceleration: the exponential increase in death rates slows down at high ages, a feature typically noticeable after about age 80  \citep{Thatcheretal:1998}. Several extensions of the Gompertz--Makeham model have been proposed to capture this phenomenon \citep{Thatcher:1999, Bebbington:2011}, the most prominent being a model where the exponential Gompertz component in \Cref{GM-haz.eq}  is replaced by a logistic function. In its most general form {the} hazard function is 
\begin{align}\label{T-haz.eq}
  h(t) = \lambda + \dfrac{A e^{\beta t/\sigma}}{1 + B  e^{\beta t/\sigma}}, \qquad A > 0, B \geq 0.
\end{align}
Several special models can be obtained from \Cref{T-haz.eq}, including the Gompertz, Gompertz--Makeham and gamma-Gompertz--Makeham models; in the last, individuals share a Gompertz component with the same age-increase, i.e., the same value of $\beta/\sigma$, but hazard levels differ between individuals. If these are modelled by a gamma-distributed multiplicative random effect, the resulting marginal hazard is a logistic function  \citep{Vaupeletal:1979, Beard:1971}.  A logistic hazard, \Cref{T-haz.eq}, increases exponentially for small $t$, but ultimately reaches a plateau of $\lambda+A/B$ as $t\to\infty$ if $B>0$.

\subsection{Extreme value models}\label{sec:EVM}

\subsubsection{Threshold exceedances}\label{GPD.sect}

 The presence or not of a cap on longevity can be viewed through the lens of extreme value statistics. The stochastic behavior of maxima and related quantities is  well-understood and is described from different viewpoints by \citet{Embrechts.Kluppelberg.Mikosch:1997}, \citet{Coles:2001}, \citet{Beirlant.etal:2004}, \citet{deHaan.Ferreira:2006} and \citet{Resnick:2006}.

A flexible approach to modeling high values of a random variable $X$ with continuous distribution function $F$ is to consider its exceedances $X-u$ above a threshold $u$ \citep{pickands:1975}. 
If, as $u$ increases to the upper support point $t_F$ of $X$, there exists a positive function $a_u$ of $u$ such that the rescaled exceedances converge to a non-degenerate distribution, then
\begin{align}\label{gpd.eq}
\Pr\left\{(X-u)/ a_u>t\mid X>u\right\} \to 1-G(t) =  
\begin{cases} (1+\xi t/\sigma)^{-1/\xi}_+, &\xi\neq 0,\\ 
\exp(-t/\sigma),&\xi=0,
\end{cases}
\end{align}
where $c_+=\max(c,0)$ for a real number $c$.
Hence the generalized Pareto distribution $G(t)$ provides a suitable statistical model for exceedances $T=X-u$ over a high threshold $u$. This distribution has a scale parameter $\sigma$, which depends on $u$, and a shape parameter $\xi$; $T$ takes values in $(0,\infty)$ if $\xi\geq 0$ and in $(0,-\sigma/\xi)$ if $\xi<0$.  Its hazard function, $(\sigma+\xi t)^{-1}_+$, is constant for $\xi=0$, declines slowly if $\xi>0$, and increases without limit as $t\to t_F$ if $\xi<0$.   With threshold $u$ and negative shape parameter, the upper limit for human lifetimes would therefore be $\psi=u-\sigma/\xi$ under this model.   A generalized Pareto variable with scale parameter $\sigma$ is threshold-stable: for $v>0$, $T-v$ is also generalized Pareto with scale parameter $\sigma_v=\sigma+\xi v$, if this is positive; the shape parameter is unchanged.  This result is useful in choosing the threshold $u$, since it gives stability relations that should be satisfied if the generalized Pareto model is adequate above $u$. Statistical aspects of this model are discussed by \citet{davison+s:1990}.

The limiting generalized Pareto survivor function,~\Cref{gpd.eq}, should provide a good approximation for data above a sufficiently high threshold $u$, so a key element in applications is the choice of $u$.  Taking $u$ too low risks the introduction of bias because the model provides a poor approximation to the distribution of exceedances, whereas taking $u$ too high may winnow the sample so much that subsequent inference becomes too uncertain to be informative. Many approaches to choosing $u$ have been proposed, including formal methods such as tests of fit and informal methods such as graphical assessment of whether the stability relations are satisfied \citep{Scarrott.MacDonald:2012,Wadsworth:2016,Bader.Yan.Zhang:2018}.

In \Cref{section4} we consider Bayesian analyses that entail specification of prior distributions. The use of objective priors is particularly important when studying the endpoint of the generalized Pareto distribution, and we apply the maximal data information prior \citep{Zellner:1977} which maximises the information from the data relative to that from the prior. 
For the exponential and  generalized Pareto distributions this reduces to the priors $\sigma^{-1}$ and $\pi(\sigma, \xi) \propto \sigma^{-1}\exp(-\xi-1)$ respectively. The posterior density for the latter is improper unless the prior is truncated; we follow  \cite{Northrop.Attalides:2016} and restrict the range to $\xi \geq -1$.  \cite{Moala:2018} discuss the validity of the prior for the Gompertz distribution, $\pi(\sigma, \beta) \propto \sigma^{-1}\exp\{\exp(1/\beta)\int_{1/\beta}^\infty \exp(-t)/t \mathrm{d} t\}$,  for a different parametrization.  In these low-dimensional settings we can use the ratio-of-uniforms method \citep{Wakefield:1991,rust} to sample from the posterior distribution.

\subsubsection{Penultimate approximation}\label{sec:penultimate}

The shape parameter determines the behavior of the upper tail of the distribution of threshold exceedances, and under mild conditions this is determined by the behavior of the reciprocal hazard function $r(t) = 1/h(t)$.  If this function has a continuous derivative $r'$ and if we write $\xi_u=r'(u)$, then $\xi = \lim_{u\to t_F} \xi_u$, but it can be shown that for $u<t_F$ the generalized Pareto distribution with parameter $\xi_u$ provides a better, so-called penultimate, approximation to the threshold exceedances \citep{Smith:1987}. The Gompertz--Makeham model, for example, has $t_F=\infty$ and 
\begin{align}%\label{GM-haz.eq}
\xi_u = -\dfrac{\beta e^{-\beta u/\sigma}}{\left( 1 + \lambda\sigma e^{-\beta u/\sigma}\right)^2},\quad u>0, 
\end{align}
so $\xi_u\to\xi= 0$ fairly rapidly as $u\to \infty$.  Thus we would expect threshold exceedances from~\Cref{GM-haz.eq} for a sufficiently high $u$ to be well-approximated by an exponential distribution, stemming from~\Cref{gpd.eq} with $\xi=0$, but for somewhat lower thresholds a better approximation would be given by $\xi_u<0$, yielding a model with a finite endpoint $\psi$.  Hence if the Gompertz--Makeham model was appropriate at all ages but a generalized Pareto distribution was fitted to exceedances of $u$, then as $u$ increases we would expect to find that the estimates of $\xi$ are initially negative but then approach zero.  

One way to assess the penultimate behavior of $F$ is to fit a model proposed by \citet{Northrop.Coleman:2014} in which generalized Pareto models with shape parameters $\xi_1, \ldots, \xi_K$ are fitted to data falling between thresholds $u_1 < \cdots < u_K<u_{K+1}=\infty$.  In order that the resulting density be smooth, the scale parameters $\sigma_1, \ldots,  \sigma_K$ over the successive intervals should satisfy $ \sigma_k = \sigma_1 + \sum_{k'=1}^{k-1}\xi_{k'}(u_{k'+1}-u_{k'})$ for $k=2, \ldots, K$.  Thus the model has parameters $\sigma_1,\xi_1, \ldots,  \xi_K$, which are estimated by treating the observations as independent, with those in the interval $(u_k,u_{k+1}]$ modelled using the truncated generalized Pareto distribution with shape $\xi_k$ and scale $\sigma_k$.  
% Although this model was proposed for choosing the threshold by finding the smallest $u_k$ above which the shape parameters are not significantly different, it can also be used to approximate $\xi_u$ by fitting a step function.

\subsubsection{Extended models}\label{sec:extGPD}

The generalized Pareto approximation for threshold exceedances holds under rather mild conditions on the underlying distribution $F$, but it may be useful to extend it by adding further parameters, in the hope of obtaining better fits in finite samples \citep{Papastathopoulos.Tawn:2013}.   
%For example,  the extended generalized Pareto model with survival function
%\begin{align}\label{egpd.eq}
%\F(t) = \left\{1+\xi \dfrac{\exp(\beta t/\sigma)-1}{\beta} \right\}_{+}^{-1/\xi}, \quad t>0, \quad \beta,\sigma>0, \xi\in\mathbb{R}, 
%\end{align}
%becomes generalized Pareto when $\beta\to 0$, gives the Gompertz distribution when $\xi\to 0$, and the exponential distribution when $\xi=\beta=0$.  Such models may be hard to fit.

Certain models simplify when certain parameters take boundary values.  For example, comparison between the Gompertz and exponential distributions based on \Cref{eq:Gompertzhazard} amounts to setting $\beta=0$, which lies on the boundary of the parameter space, so standard large-sample results for likelihood ratio tests do not apply; the asymptotic significance level for a positive likelihood ratio statistic $w_{\rm Gom}$ is $\frac{1}{2}\Pr(\chi^2_1>w_{\rm Gom})$, one-half of its usual value, and the significance level for $w_{\rm Gom}=0$ is unity.   Limiting distributions such as these can give poor finite-sample approximations, so simulation from the fitted null model is generally preferable for model comparison. \Cref{tab:bootlrt} illustrates this for the England \&~Wales semi-supercentenarian data discussed in \Cref{sec:ew_data_analysis}.  The $p$-values obtained by simulating $9999$ datasets from the fitted exponential model with the same truncation bounds are systematically larger than the asymptotic $p$-values, because the asymptotic probability that $w_{\rm Gom}=0$ can be far from its finite-sample value.

\begin{table}

\caption{Likelihood ratio tests comparing Gompertz and exponential models for the England \&~Wales data over a range of thresholds $u$, with the number of exceedances $n_u$, the likelihood ratio statistics $w_{\rm Gom}$ the bootstrap $p$-values $p_{\text{b}}$ and asymptotic $\frac{1}{2}\chi^2_1$ $p$-values $p_{\text{a}}$ for the null hypothesis {$\beta=0$}.  Also shown are the corresponding likelihood ratio statistics $w_{\rm GPD}$  for comparison of the generalized Pareto  and exponential models,  with their asymptotic significance levels $p_{\rm GPD}$.}
\label{tab:bootlrt}
\begin{center}
\begin{tabular}{crccccc}
\toprule
$u$ & $n_u$ & $w_{\textrm{Gom}}$ & $p_{\text{b}}$ & $p_{\text{a}}$ & 
$w_{\textrm{GPD}}$ & $p_{\textrm{GPD}}$\\
\midrule
108 & 510 & 7.98 & 0.004 & 0.002 & 7.43 & 0.006\\
109 & 225 & 0.98 & 0.198 & 0.161 & 1.08 & 0.298\\
110 & 85 & 2.70 & 0.074 & 0.050 & 2.71 & 0.099\\
111 & 39 & 0.81 & 0.274 & 0.184 & 0.81 & 0.368\\
\bottomrule
\end{tabular}
\end{center}
\end{table}

\section{DATA ANALYSES} \label{section4}

\subsection{Netherlands} \label{subsection:Netherlands}

We first demonstrate the presence of penultimate effects that could result in underestimation of the lifespan due to unreliable extrapolation. The data, from \cite{Einmahl:2019},  concern individuals who died aged at least 92 years between 1986 and 2015 and are from a Dutch population register; although unvalidated they should be free of age-ascertainment bias.  There are lifetimes in days and the months and years of birth and death for 304~917 individuals, plus 226 persons whose lifetimes are unknown and so are interval-censored. All of these lifetimes are interval-truncated, and we treat them as such in our analysis.  There is no sign of gender or cohort effects from age 105 onwards.

\begin{figure}[t] 
 \centering 
 \includegraphics[width=\linewidth]{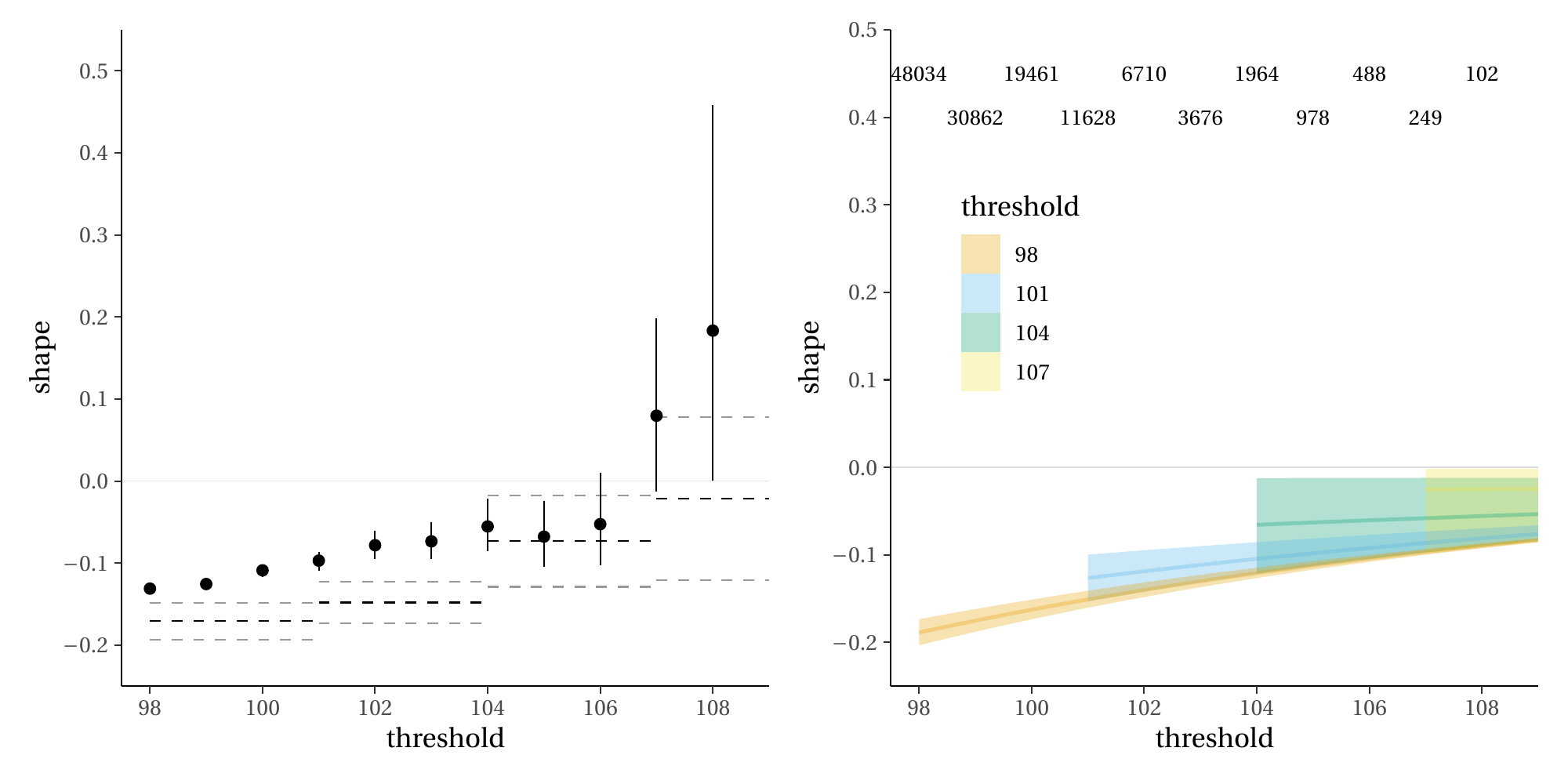}
 \caption{
Penultimate effects in the Dutch data. Left: maximum likelihood estimates of the shape parameter with thresholds $98, \ldots,  108$ years (dots) and 95\% profile likelihood ratio confidence intervals and Northrop--Coleman estimates (dashed) with thresholds $98$, $101$, $104$, $107$ and 95\% confidence intervals (dashed gray). Right: estimated penultimate shape parameter $\xi_u$ (thick line) for the Gompertz model fitted to exceedances over thresholds $98$, $101$, $104$, $107$ years, with  95\% pointwise confidence regions (shaded). The numbers of exceedances above the yearly thresholds  are reported in the right-hand panel.
 }
 \label{Dutch.fig}
 \end{figure}

If the generalized Pareto distribution were a good approximation to lifetimes above age $u$, then estimates of its shape parameter $\xi$ would be roughly constant for fits  using higher thresholds. The left-hand panel of \Cref{Dutch.fig}, however, shows that these estimates increase steadily as $u$ increases, suggesting that the lowest reasonable threshold here is 102 years.
 
For a deeper analysis we fit the \cite{Northrop.Coleman:2014} piecewise generalized Pareto model, handling truncation and censoring as described in \Cref{lik-contribs.eq}. With $K=3$ and thresholds $u_1=98$, $u_2=101$, $u_3=104$ and $u_4=107$ years, the estimated shape parameters show a steady increase which is compatible with penultimate effects.
% ; the estimates and standard errors are $\hat\sigma_1=2.54_{0.021}$, $\hat\xi_1= -0.171_{0.011}$, $\hat\xi_2= -0.148_{0.014}$,  $\hat\xi_3=-0.073_{0.028}$ and $\hat\xi_4= -0.022_{0.057}$.  
 This partially explains the results of \cite{Einmahl:2019}, whose fits are dominated by the lower ages. 

The piecewise model can be used to check the hypothesis of equal shape through likelihood ratio tests:  if $\xi_k = \cdots =\xi_K$, then the model is generalized Pareto above $u_k$ with parameters $(\sigma_k, \xi_k)$. Comparison of the piecewise model with thresholds $u_1, \ldots, u_{4}$ and a generalized Pareto model above $u_k$ leads to rejection of the latter  at thresholds 98 and 101.  For threshold $104$ the $p$-value is $0.49$, so the two models appear to fit equally well.

We can also fit the Gompertz model to exceedances over a range of thresholds.  The maximised log-likelihoods for the Gompertz  and generalized Pareto models are equivalent from threshold 102 onwards and the models seem very similar until threshold 106. However, the Gompertz model cannot capture a decreasing hazard function; it reduces to an exponential model above high thresholds, which seems implausible based on the generalized Pareto fits at thresholds 107 and 108. 
If the Gompertz model were adequate, we would expect its tail behavior, given by the penultimate shape $\xi_u$ obtained from fitted models, to broadly agree with the generalized Pareto shape estimates. 
The right-hand panel of \Cref{Dutch.fig} shows 95\% pointwise intervals for the  curves corresponding to the penultimate shape  $\xi_u$ of the fitted Gompertz models  at thresholds $98, 101, 104$ and $107$. These curves agree with the generalized Pareto shape estimates in the range 104--107 years, but not elsewhere; in particular, the shape parameter estimates at 107 seem to be positive, but those for the Gompertz model are capped at zero.   The Gompertz model does not seem to be flexible enough to capture the penultimate properties of the data. 

Validation of these data might change the results, but it appears that penultimate effects can be detected in these data until age 104 years.  Although the reduced sample sizes render them undetectable, it seems plausible that such effects persist above age 104.  Extrapolation is therefore unreliable if based on a generalized Pareto fit to individuals aged below 105 years and, if possible, a higher threshold is to be preferred.  The inflexibility of the Gompertz hazard function suggests that it should be avoided for fits at very high ages. 

\subsection{Japan}
\label{subsec:japanese}

Although we have stressed the central role of interval truncation, other types of {data coarsening} arise.  We illustrate the effect of interval censoring and the asymmetric nature of inferences on the lifespan using data from \cite{Hanayama/Sibuya:2016}, who analyze annual death counts since 1947 for centenarians using information from the Japanese Annual Vital
Statistics Report (see \url{https://www.mortality.org/}).  There are 10~440 semi-supercentenarians in a total of 122~719 unvalidated records.   \citeauthor{Hanayama/Sibuya:2016} use a multinomial likelihood for the counts in each combination of age at death and birth cohort, based on an underlying generalized Pareto model,  to compare the accumulated damage and programmed ageing theories. They find better agreement with the latter, their lifespan estimate based on a subset of the women being 123 years.  Although they only consider birth cohorts for which no death has been reported for three consecutive years, up to and including 1898 for males and 1894 for females, the paper includes counts until 1898 for women, for whom $c_2=2014$. This selection mechanism is analogous to the hidden truncation scheme described in \Cref{sec:samplingscheme}; the lifetimes are interval-censored and right-truncated.   

\begin{figure}[t]
\centering 
\includegraphics[width = \linewidth]{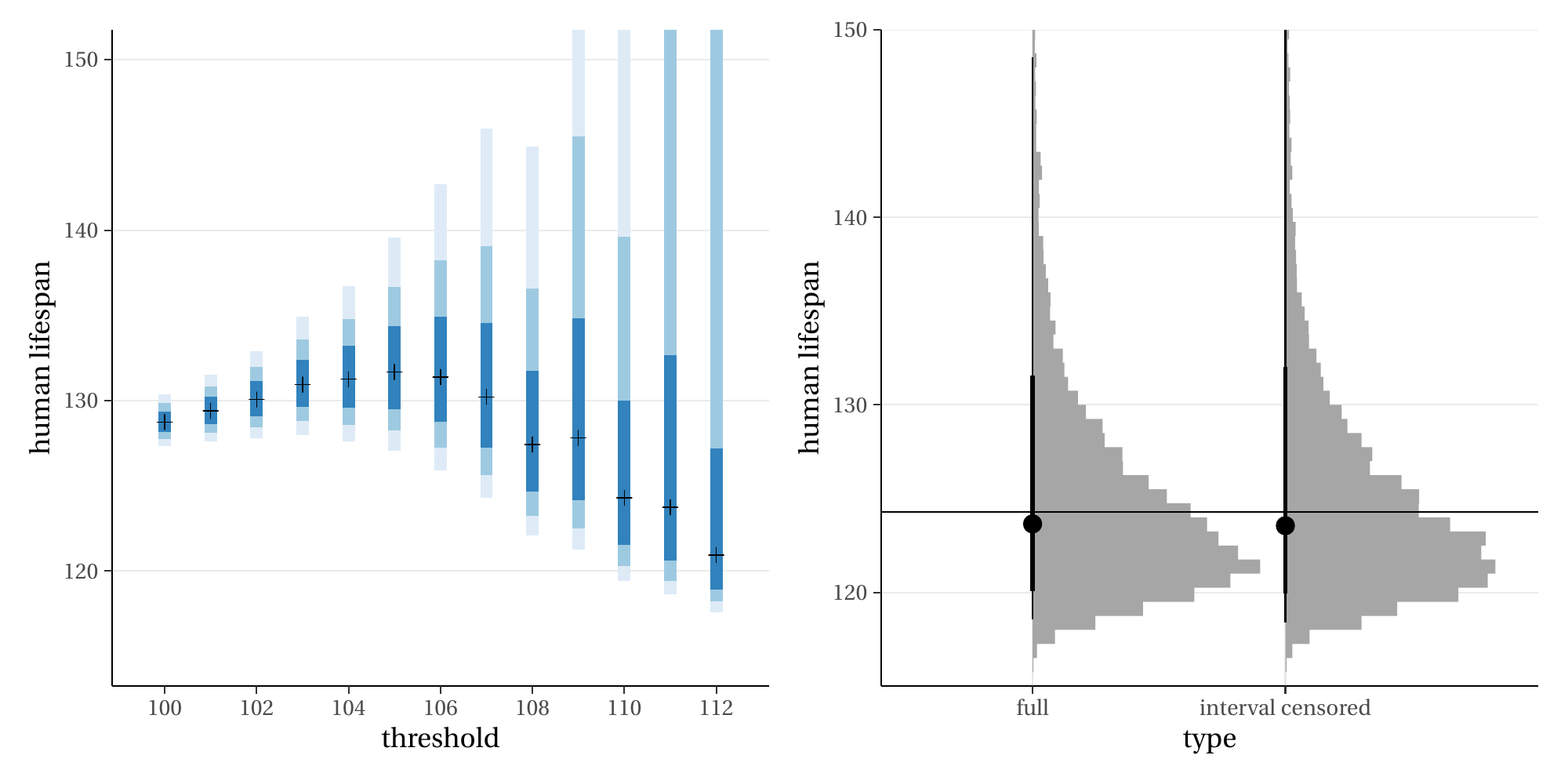}
 \caption{Analysis of Japanese data.  
  Left: maximum likelihood estimates (crosses) with  50\%, 75\% and 90\% profile likelihood confidence intervals for the endpoint $\psi$ for a range of thresholds; some intervals end above 150 years.  Right: histograms of the sampling distributions for maximum likelihood estimators of $\psi$ based on fully-observed (left) and interval-censored (right) simulated datasets.  The panel shows the median (circle), 75\% and 95\% sampling limits, and the horizontal black line is the endpoint $\psi=124.3$ of the simulation distribution. Less than 3\% of the endpoint estimates exceed 150 years, including those for the roughly 0.3\% of the shape estimates that are positive.}
 \label{fig:japanese}
\end{figure}

Point estimates of the human lifespan do not paint the whole picture.  We compute profile likelihood confidence intervals  for the endpoint $\psi=u-\sigma_u/\xi$ at levels 50\%, 75\% and 90\%. The maximum likelihood estimates for $\xi$ are negative for all thresholds $u$ in the range 100, 101, \ldots, 116 years and this translates into finite point estimates $\hat\psi$, but for higher thresholds the confidence intervals are strongly asymmetric, suggestive of a very large upper limit; see the left-hand panel of \Cref{fig:japanese}. As the threshold increases, $\hat\psi$ rises and then drops back, but its uncertainty increases because of the smaller sample size.  

Tabulating death counts rather than publishing individual records increases data privacy but could lead a loss of precision. To assess this, we computed the maximum likelihood estimates $\widehat{\psi}$ using both the full data and interval censoring to mimic tabulation for $10\ 000$ datasets of size $n=513$ with excess lifetimes simulated from the generalized Pareto model  with $\sigma=1.546$ and $ \xi=-0.108$, the estimates for exceedances over $u=110$; like the original data, the simulated lifetimes were right-truncated. The histograms of the estimates shown in the right-hand panel of \Cref{fig:japanese} display downward bias, and suggest that conventional symmetric confidence intervals centered on the estimated endpoint will be extremely poor.    The sampling distributions are very similar, so the loss of information due to interval censoring is small, though it depends on the width of the age bins.  This experiment suggests that this loss and any bias are immaterial for yearly bins, at least if the shape parameter is negative.

This analysis is consistent with that in \Cref{subsection:Netherlands} in suggesting that conclusions from data below around 103 years are too unstable to be useful, and it underscores the asymmetry of uncertainty for the lifespan $\psi$.  Tabulation by age at death and birth cohort does not seem to increase uncertainty substantially, but it does preclude validation of individual records and therefore should be avoided. 

\subsection{England \&~Wales} \label{sec:ew_data_analysis}

In this section we illustrate the complications that truncation causes for Q-Q plotting and the radical reduction in uncertainty when extremal models are restricted.  We use data on semi-supercentenarians in England \&~Wales who died between 2000 and 2014 \citep{ONS:2016}, provided by the UK Office for National Statistics (ONS). All the lifetimes are interval-truncated and the maximum age achieved is just short of 115 years.   All records for men and for lifetimes above 109 years were validated, but only a stratified sample of the remaining records was checked. There are 3624 females and  318 males. Recent estimates \citep{ONS:2020} conclude that the numbers of male and female semi-supercentenarians have respectively increased by about 100\% and 50\% over the last decade, so the imbalance is diminishing.   

\begin{figure}[thbp]
\centering 
\includegraphics[width = \linewidth]{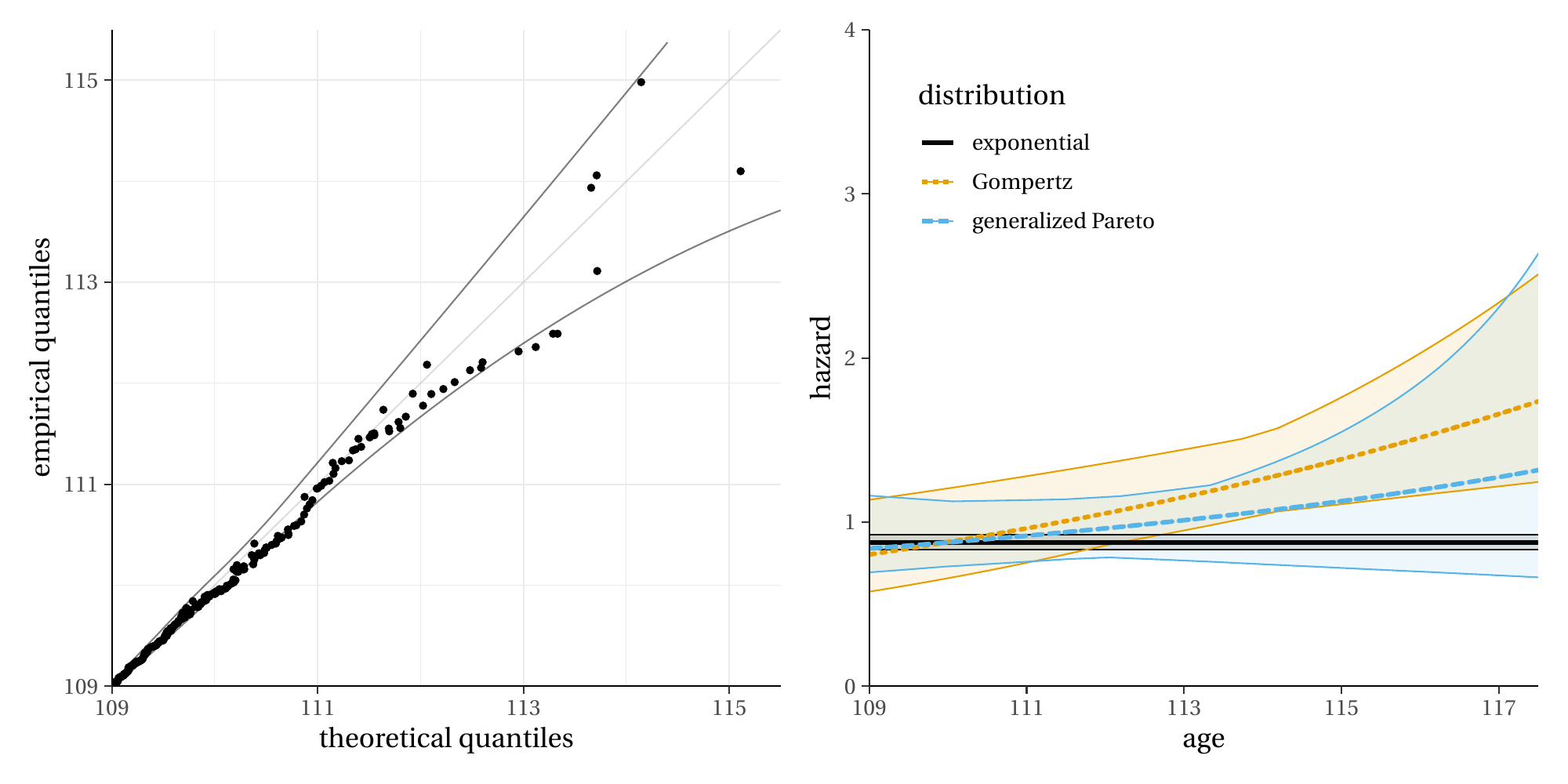}
 \caption{Analysis of exceedances above age 109 for England \&~Wales semi-supercentenarians. Left: 
exponential Q-Q plot accounting for truncation, with approximate 90\% pointwise confidence intervals constructed as described in \Cref{subsec:diagplot} and in \Cref{sec:appendixD}. Right: Bayesian analysis for Gompertz, generalized Pareto  and exponential hazard functions using maximal data information priors. The plot shows posterior median (lines) and 50\% posterior predictive intervals (shaded) for the hazards.}
 \label{fig:englandwales}
\end{figure}

We fitted Gompertz and generalized Pareto models for a range of thresholds.  Likelihood ratio tests show no significant gender effects above 106 years.  The generalized Pareto fits show strong evidence of negative shape parameters for thresholds 105--107.  The exponential model is a much worse fit than the Gompertz and generalized Pareto models until threshold 109; see \Cref{tab:bootlrt}. The Q-Q plot in the left-hand panel of \Cref{fig:englandwales} compares these data with the exponential model for  threshold 109 years.  The effect of the truncation is that the plot is non-monotone.  Lifetimes between 110 and 111 seem too small compared to the fitted model.

Adoption of an over-simplified model may lead to under-estimation of the model uncertainty. To illustrate this we perform a Bayesian analysis, using the priors and sampling approach described at the end of \Cref{GPD.sect} to sample from the posterior densities for the generalized Pareto, exponential and Gompertz models. We computed 50\% intervals using the half-depth method \citep{ggdist}, and display them in the right-hand panel of \Cref{fig:englandwales}. As the exponential hazard is constant, the intervals are very narrow. The median Gompertz and generalized Pareto hazards functions are increasing, but uncertainty for the latter encompasses constant or decreasing hazard; the latter is impossible for the Gompertz model, which is less uncertain overall.  
%When the shape parameter is negative, the generalized Pareto hazards rise to infinity at the finite endpoint and thereafter are undefined.   

Simplifying inferences by choosing the exponential or the Gompertz models risks a large understatement of uncertainty, particularly for the former. 

\subsection{Italy and France}
\label{subsection:FranceandISTAT}

The Italian National Statistics Institute \Istat{}\ has recently produced a validated database with 3373 women and 463 men in Italy who were at least 105 years of age at some point from 1 January 2009 to 31 December 2015; it covers birth cohorts from 1896 to 1910, and 953 individuals were alive at the end of 2015.  The data, which are left-truncated and right-censored, were analyzed by \citet{Barbi:2018}, who fitted a Gompertz model and found a small but statistically significant cohort effect, a small and statistically insignificant effect of gender and a plateauing of the hazard function after age 105, corresponding to an exponential excess lifetime distribution.  

New French data added to the \IDL{} in 2019 consist of 9612 semi-supercentenarians and 241 supercentenarians; all of the latter but only some of the former were validated. These data are interval-truncated. Analysis of the Italian data and of these new French data using a variety of models led \citet{Belzile:2020}  to conclude that after age 108 the excess lifetime are compatible with an exponential distribution, with no evidence of birth cohort effects.  The only significant gender effect they found was for the French men, whose mean estimated survival time after age 108 was 0.9 years, with the corresponding value for the Italian data and for the French women being 1.43 years; the respective 95\% confidence intervals are $(0.70, 1.10)$ and $(1.32,1.54)$ years.   The estimated one-year survival probabilities in the French data  based on an exponential distribution are 0.50 for women, with 95\% confidence interval (0.46, 0.53), and 0.33 for men, with 95\% confidence interval (0.24, 0.41).  The reason for this difference in the French data alone is unclear; it may be a false positive due to repeated significance testing, there may be some issue with the data, or it may be a real effect whose cause is unknown.

\cite{Belzile:2020} also use these datasets to study the power of likelihood ratio tests for detecting a finite lifespan: pooling data from French, \Istat{} and \IDLold{}, they find that such tests would have power of over 80\% for detecting that the endpoint of a generalized Pareto distribution lies below 130 years and of over 50\% for it to lie below 140 years.
\subsection{International Database on Longevity} 
\label{subsect: IDL-gender-countries}

The analyses by  \cite{Gampe:2010} and \cite{Rootzen.Zholud:2017} of the 507 supercentenarians in the 2016 version of the \IDL{} led to the conclusion that survival above age 110 was the same for women and men, for different countries, and for persons born earlier or later, and that these lifetimes could be described by an exponential distribution with mean 1.42 years, with 95\% confidence interval $(1.28,1.56)$ years. The 2021 version of the \IDL{} \citep{IDL:2021} includes about twice as many supercentenarians as the 2016 version, and the countries included have changed somewhat. \Cref{sec:appendixF}  contains updated versions of Tables 1--5 of \cite{Rootzen.Zholud:2017}.  The general conclusions are the same: there are no detectable differences in survival between women and men or between countries. The $p$-value for testing for differences in the USA between persons who died earlier and later is small, but differences in Europe go in the other direction and for the joint data there is no indication of an effect. An increase over time in survival past age 110 would be a striking finding, but we have made no adjustment for multiple testing, and such differences could be due to chance or perhaps to changes in validation procedures. Hence we again conclude that survival after age 110 can be summarized by an exponential distribution. For the new data the exponential mean was estimated to be 1.38, with 95\% confidence interval $(1.29,1.48)$, so the estimate lies close to the centre of the previous 95\% confidence interval. The new estimated probability of surviving one more year is 0.48, with 95\% confidence interval $(0.46, 0.51)$.

\cite{Belzile:2020} and Sections~\ref{subsection:Netherlands}--\ref{subsection:FranceandISTAT} of the present paper study data on semi-supercentenarians and supercentenarians from Italy, France, the Netherlands, Japan and England \&~Wales. For each of these countries the analysis indicates that  the hazard function appears to have plateaued by age 109, but the large uncertainty does not rule out a decreasing hazard beyond this point.

 The 2021 \IDL{} includes  both semi-supercentenarians and supercentenarians for Austria, Belgium, Quebec, Denmark,  Germany,  England \&~Wales and France. \Cref{table:thresholds} in \Cref{sec:appendixF} indicates that plateauing has also occurred  well before age 109 for  these countries, for which the  estimated yearly probability of survival after age 109 is  0.45, with 95\% confidence interval $(0.36, 0.53)$. 

\section{REVIEW OF ANALYSIS STRATEGIES}
\label{sec:strategy}
{Although writing $t_F=\sup\{t: F(t)<1\} = \infty $ is natural for mathematicians,} a lifetime distribution with unbounded support typically meets with skepticism. Infinity evokes the echo of immortality, but every observed lifetime has been and always will be finite, so careful translation of mathematical truths into everyday language is required. Unbounded support does not imply infinitely long lifetimes (immortality),  and the fact that a certain age $\breve t$ is highly unlikely to be surpassed under a fitted model does not imply that $t_F \leq \breve t < \infty$. 

Past strategies  to assess whether $t_F < \infty$, corresponding to a cap, have been direct or indirect. In this section we review some of the more prominent of them.

\subsection{Tracing the record}
\label{sec:record}

A superficially natural line of reasoning to investigate limits to lifetimes is to track the longest recorded lifetimes, either globally or per birth cohort and sometimes also per calendar year. Prompted by an early postulate that \textsl{`the length of life is fixed'} \citep{fries1980} since \textsl{`there has been no detectable change in the number of people living longer than 100 years or in the maximum age of persons dying in a given year'} (p.~130, ibid.) record lifetimes were consulted for analysis. Both claims have long been disproved, by newer and better data (\citealp{RobineCaselli:2005,wilmoth:2000}), but maximum recorded lifetimes have remained an initial and also often the final point in subsequent reasoning. Argument from the times between records is problematic both because the population size is not stable and because these times are highly variable even in a stationary setting.

Fitting trend lines to annual maximum lifetimes likewise says little to nothing about $t_F$, especially if the size of the population from which the maximum is derived is not taken into account, and it says even less if the observation scheme is ignored. \cite{dongetal2016}, the paramount example of this approach, has been shown to be flawed \citep[e.g.,][]{Rootzen.Zholud:2017,Keiding:2018,IDL:2021} despite a positive review by \emph{Nature}. 

Analyzing trends in maxima derived from subsets of observations and attempting inference from this sample of maxima is reminiscent of the block maximum method of extreme value theory, though it does not use this fully. Moreover, block maxima must be handled with care because of the increasing number of supercentenarians over time and the truncation and censoring of the underlying lifetimes;  threshold exceedance approaches (\Cref{GPD.sect}) are preferable.

\subsection{Hazard trajectories at high ages}
\label{sec:hazardtraj}

Another indirect line of reasoning involves hazard trajectories at the highest ages. \cite{Barbi:2018} derive a plateau of constant mortality after age 105 for the data discussed in Section 4.5, implying an exponential distribution, which has unbounded support, `underwriting doubt that any limit is yet in view' \citep[p.~1641]{Barbi:2018}. A constant hazard function excludes finite $t_F$, since for bounded support the hazard would have to increase to infinity sufficiently quickly (\Cref{sect.survival-basics}). A continuously increasing hazard such as that of the Gompertz model has been interpreted as indicating a finite lifespan \citep{Gavrilova:2020}.  This is incorrect: the Gompertz, Gompertz--Makeham and many other distributions with hazards increasing to infinity have unbounded support.  Finding that the Gompertz hazard holds for advanced ages would be a notable result, but even if true it would not imply that $t_F$ is finite.

\subsection{Extreme value analysis}
\label{sec:EVstats}

Extreme value statistics, and specifically the modeling of threshold exceedances, directly addresses the question of an upper limit to the lifetime distribution. Section 3.4 presented the underlying theory and Section 4 gives some empirical results and discusses practical intricacies. An early study of this type was~\cite{Aarssen-dehaan1994}, and more recently \cite{Gbari-etal:2017}, \cite{Hanayama/Sibuya:2016} and \cite{Einmahl:2019} {have} employed this strategy. The data of the latter two papers were reanalyzed in Sections 4.1 and 4.2. \cite{Gbari-etal:2017} fit generalized Pareto distributions to lifetimes from the Belgian national population register for the 1886--1904 birth cohorts. The study includes only extinct cohorts, so the considerations of Section 3.2 apply. They discuss different methods for  threshold choice and decide on  thresholds of approximately 99 years for men and 101 years for women. As a result they get negative estimated shape parameters and finite and reasonable estimates for the maximum life span.  Q-Q plots indicate good fit for females, and perhaps a bit less so for males, but estimates are largely determined by the bulk of the data, at ages 105 and lower.  The distorting effect of grouping the observed lifetimes by year of death rather than year of birth, as done in \cite{Einmahl:2019}, was addressed in Section 3.2.  In periods of increasing numbers of oldest-old, as currently seen in many countries, this uneven influx of observations at the `young' end of the samples biases the lifetime distribution towards the lower ages.

\section{CONCLUSIONS}

Many, perhaps most, statistical studies of extreme human lifetimes are problematic. 

A first class of problems concerns the data, which are validated with insufficient care, or sometimes not validated at all.  Lifespan estimation is particularly vulnerable to poor data, so punctilious verification of individual extreme lifetimes is essential.   Another issue is the combination of datasets that bear on different phenomena and thus should not be analyzed together, and yet another is the difficulty in replicating earlier studies, owing to lack of access to data, apparently arbitrary or ill-documented decisions about data-handling, or a failure to provide the code to reproduce an analysis.

A second class of problems is that the observation scheme is unknown or not taken into account.  Data culled from news reports, social media and the like are opportunity samples, so  may be highly biased and cannot form the basis of reliable inference, whereas failure to correct for the truncation and censoring present in most observation schemes will lead to biased estimation  of the human lifetime distribution.

A third class of problems concerns extrapolation beyond the largest lifetimes yet observed to the human lifespan.  The force of mortality increases up to age 100 or more, and there has been confusion over whether this increase implies that there is a finite human lifespan; this may be inferred only if the force of mortality rises indefinitely before some finite time.  The natural stochastic framework for modeling rare events is extreme value theory, which involves extrapolation from asymptotic models fitted to data at finite levels.  Articles using extreme value theory have tended to include data for which the limiting models are inapplicable, in some cases despite an apparently good fit due to ignoring the observation scheme.   Extreme-extreme valuevalue fits to lifetimes below 105 years or even more are unstable, suggesting that the asymptotic regime has not yet been reached, and it seems plausible that only  lifetimes exceeding a threshold of 109 years or so lie in the asymptotic domain.  This dramatically reduces the size of the relevant sample, which roughly halves for each one-year increase in the threshold in this range.

The statistical analysis of lifetime data is full of traps for the unwary, and the widespread and natural interest in the limits to human life have led many investigators to tumble unwittingly into them. We hope that this review will lead to better studies than some past ones, even those published in supposedly highly prestigious journals.

\begin{figure}[thbp]
\centering 
\includegraphics[width = \linewidth]{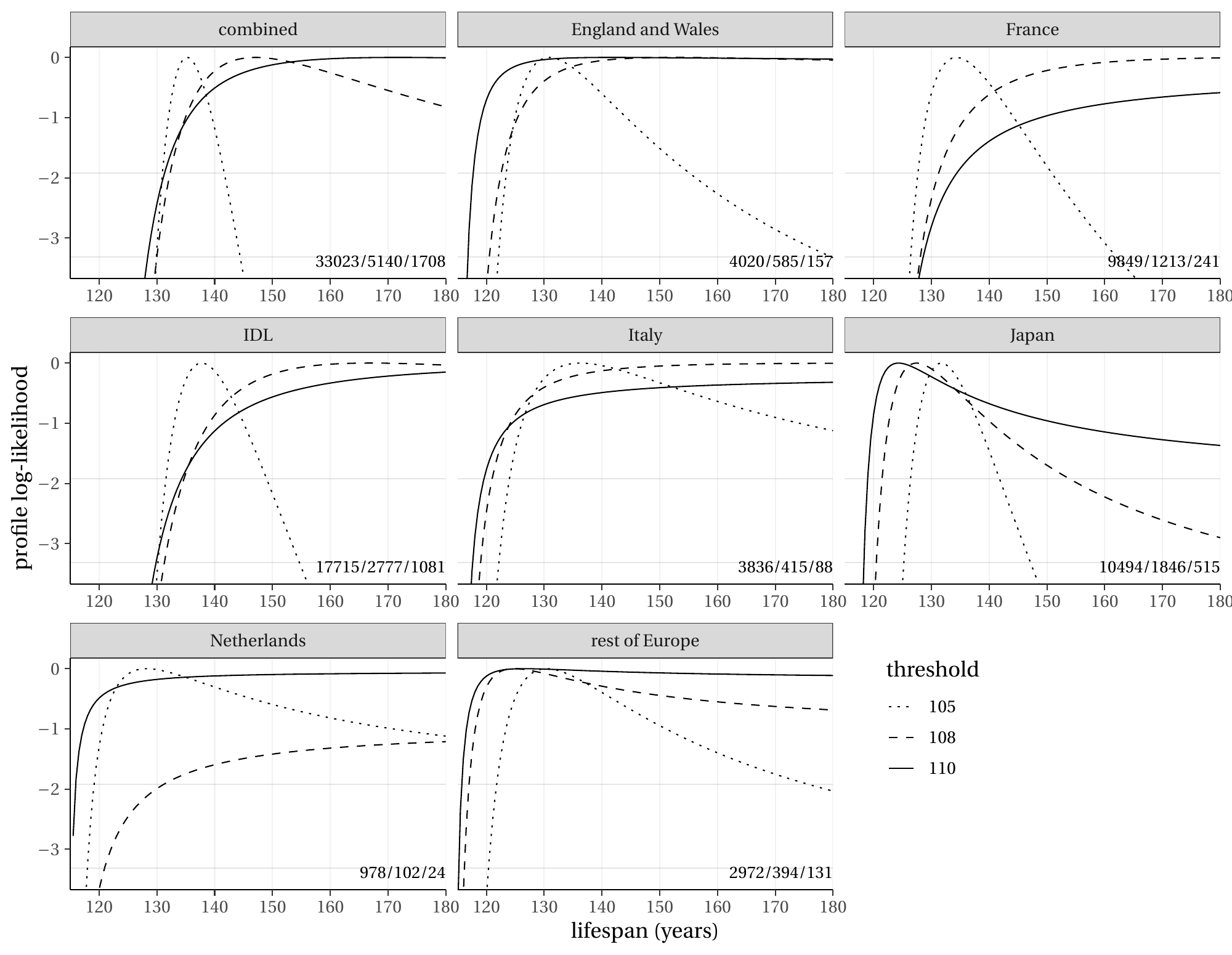}
 \caption{Profile log likelihoods for the human lifespan (years) based on generalized Pareto distributions fitted to exceedances over thresholds 105 (dotted), 108 (dashed) and 110 years (solid), separately for national datasets. England \& Wales includes all supercentenarians from \IDL{} and the new ONS data; Italy is based on \Istat{} data alone; `rest of Europe' consists of \IDL{} data for Austria, Belgium, Denmark, Finland, Germany, Norway, Spain, Sweden and Switzerland;  and `\IDL{}' includes rest of Europe, England and Wales, France and \IDL{} Italy records, plus data from Quebec and the USA (excluding ages 105--109 for the latter). The combined fit includes Italy (i.e., \Istat{}), \IDL{}, Netherlands and Japan. The horizontal lines indicate limits of 95\% and 99\% confidence intervals based on a $\chi^2_1$ distribution. The counts at the bottom right of each panel give the numbers of exceedances over the three thresholds.}
 \label{fig:summary}
\end{figure}

Is there a cap on longevity?  If there is one, it appears not to depend on gender, nationality or period of death, as we saw in \Cref{subsect: IDL-gender-countries}.   Although an exponential model for excess lifetimes above age 109 is consistent with the data, neither it nor the Gompertz model allows a finite human lifespan.  \Cref{fig:summary} summarizes evidence on the human lifespan $\psi$ from the datasets considered in this paper.  Each panel shows the profile log-likelihood for $\psi$ from fits of the generalized Pareto distribution to lifetimes exceeding 105, 108 and 110 years, taking truncation and/or censoring into account.  The data for the \IDL, England \&~Wales, France and Italy were validated, but not those for Japan and the Netherlands.  Lifetimes for individuals in each dataset were treated as independent and identically distributed, but the second generalized Pareto parameter can vary, so the fits to different datasets may differ despite the common upper limit. In each panel the support of the fitted generalized Pareto model lies above the oldest lifetime for that dataset, so the combined support lies above the longest validated lifetime of 122 years and 164 days, or 44724 days, for the Frenchwoman Jeanne Calment. No individuals appear twice in the `combined' panel.

For threshold 105 years, i.e., for semi-supercentenarians and supercentenarians, the estimated lifespans  for the individual datasets lie in the range 125--140, and the 95\% confidence intervals have finite upper limits except for Italy and the Netherlands.  The combined lifespan estimate is around 135 years, with 95\% confidence interval of around $(131, 143)$ years.   For threshold 108 years, only Japan has an upper limit of less than 180 years for its 95\% confidence interval.  The other datasets put no limit to the human lifespan, and the combined estimate is around 147 years, with approximate 95\% confidence interval $(133, 180+)$.  For threshold 110 years, the only datasets appearing to give a 95\% confidence interval with a finite upper limit are Japan and, possibly, rest of Europe.  The combined estimate seems to be above 180 years, if it is finite; the 95\% confidence interval has lower bound 131 years or so and an apparently infinite upper bound.  These confidence limits presuppose that standard likelihood theory is applicable, but even if this is not the case, the profiles suggest that the human lifespan lies well beyond any lifetime yet observed or that could be observed in the absence of major medical advances.

Although other models are also compatible with the data, the exponential distribution has a convenient interpretation due to the memoryless property, as the probability that a supercentenarian will survive an extra year does not depend on his or, more usually, her, current age. This probability is estimated to be 0.50, with 95\% confidence interval $(0.46,0.53)$, so one can view this event as a coin toss.  Even if this imposes no upper bound on lifetimes, it does not imply that the highest human lifetime will be very large: the probability that a 110-year old will live a further 20 years is that of 20 successive heads when tossing a fair coin,  a near one-in-a-million event. A record age of 130 years is unlikely, even with the present increase in the number of supercentenarians.

\subsection*{Reproducibility}

The  \Rlang{} package \texttt{longevity} implements the methodology used in the paper and is available for download from \url{https://github.com/lbelzile/longevity}.  The \Rlang{} code used to produce the analyses is available from \href{https://github.com/lbelzile/arsia-longevity}{Github}. 

\href{https://github.com/OGCJN/Rejoinder-to-discussion-of-the-paper-Human-life-is-unlimited---but-short}{LATool} is a \textsf{MATLAB} toolbox for life length analysis that makes alternative analyses possible. It is available from \texttt{https://doi.org/10.1007/s10687-017-0305-5}.

{Data from the \IDL{} are freely available after registration.  The Dutch and Japanese data are available from the respective publications and via \texttt{longevity}.  For access to the Italian data, see \citet{Barbi:2018}.}

\subsection*{Acknowledgements}

We are grateful to Ngaire Coombs and the ONS for providing updated data for England \&~Wales, to the International Database on Longevity staff and volunteers, to the Swiss National Science Foundation for financial support, and to Tatiana Moavensadeh-Ghasnavi for helpful insights. 

% \bibliographystyle{ar-style1}
% \bibliography{libraryLongevity}

\appendix 

\section{Sampling frame of the International Database on Longevity} \label{sec:appendixA}
The \IDL{} data available from \url{supercentenarian.org} at the time of writing only includes records of dead individuals and thus does not match exactly the description in \cite{IDL:2021}; data for Switzerland and Italy have also been removed for confidentiality reasons. 
% The exact lifetime (in days) varies among individuals due to leap years. \cite{IDL:2021} discuss the varying sampling frame, but the tables and description do not match the database available online.

\begin{figure}[t!] 
\centering 
\includegraphics[width=\textwidth]{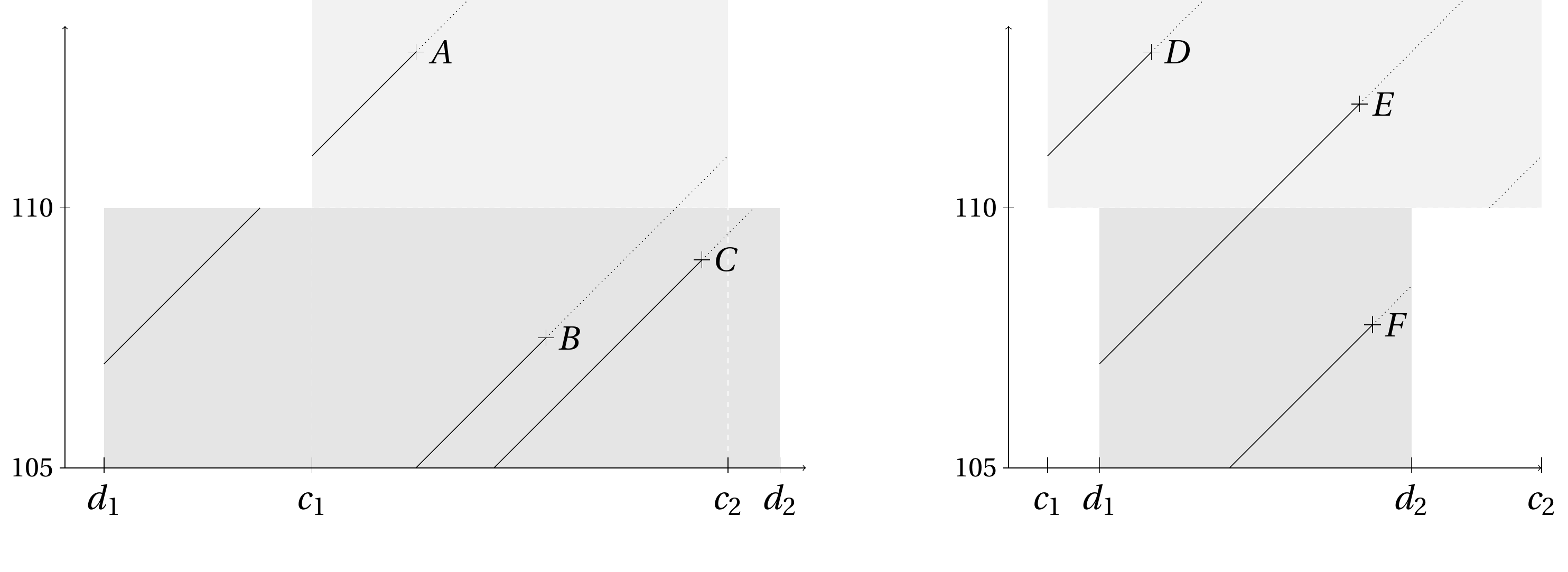}
\caption{Lexis diagrams of the \IDL{} data for countries with semi-supercentenarian records, consisting of sampling frame defined as regions in  $[105,110) \times [d_1, d_2]$ and $[110, \infty) \times [c_1, c_2]$. Left:  case $d_1 < c_1$; the lifetime trajectory of individual $A$ is observed above age 105 in $[d_1, x^{110}_A] \cup [c_1, c_2]$, that of individual $B$ in $[x_B^{105}, c_2]$, and that of individual $C$ in $[x_C^{105}, x_C^{110}]$. Right: case $c_2 > d_2$; the lifetime trajectory of individual $D$ is observed in the interval $[c_1, c_2]$, that of individual $E$ in $[d_1, c_2]$, and that of $F$ on $[x_F^{105}, d_2] \cup [x_F^{110}, c_2]$.}
\label{fig:lexisdiag2}
\end{figure}

Recent versions of the \IDL{} include both semi-supercentenarians who died between calendar dates $d_1$ and $d_2$ and supercentenarians who died between calendar dates $c_1$ and $c_2$. The sampling frame thus consists of two regions, $[d_1, d_2] \times [105,110)$ and $[c_1, c_2] \times [110, \infty)$, with potentially different truncation bounds. This complicates the analysis because many lifetimes are doubly interval-truncated and handling the constraints requires a careful case-by-case analysis. Inclusion criteria can be deduced from the Lexis diagram by considering the birth date of all individuals of a given country. All countries in the \IDL{} database with semi-supercentenarian have $c_2 > d_1$, and we assume this hereafter. Let $x^{110}$ and $x^{105}$ denote the calendar date at which an individual reaches  110 and 105 years respectively; \Cref{fig:lexisdiag2} shows a typical \IDL{} sampling frame with some scenarios discussed next.

If we were only considering supercentenarians, lifetimes would typically be left-truncated by the age reached at $\max(x^{110}, c_1)$ and right-truncated at $c_2$. However two  instances can lead to double interval-truncation. The first can occur when $d_1 < c_1$; if $x^{105} < d_1 < x^{110} < c_1$, as for trajectory $A$, the trajectory is observed in $[d_1, x^{110}] \cup [c_1, c_2]$. % which reduces to $[d_1, c_2]$ if $x^{110} > c_1$. 
The second  can occur when $d_2 < x^{110} <c_2$, as for trajectory $F$, in which case some lifetimes lie in $[\max(d_1, x^{105}), d_2] \cup [x_{110}, c_2]$.
% and else $[\max\{d_1, x^{105}\}, c_2]$ if $d_2 \geq x_{110}$.
If $d_1 < x^{110}  < d_2$, some individuals could have traversed the set $[d_1, d_2] \times [105,110)$ even when they died beyond 110 years, so their minimum observable age is that reached at $\max(x^{105}, d_1)$.  Similarly, people who died in $[105,110)$ could have become supercentenarians and seen their death recorded if their 110th anniversary $x^{110}$ lies in $[c_1, c_2]$; then their lifetime is right-truncated by the age they would have reached at $c_2$. For all other individuals in $[105,110)$, the maximum observable age is capped at that reached at $\min(d_2, x^{110})$. If $x^{110} > c_2$ and we observe a failure in $[d_1, d_2] \times [105,110)$, then the maximum possible age achieved is $\min(d_2, x^{110})$.   The left-hand panel illustrates the case when  $d_1 < c_1$, whereas the right-hand panel shows an example where $d_2 < c_2$. This example illustrates the importance of accurate metadata, the lack of which precludes correct statistical inference.

\section{Extinct cohort analysis} \label{sec:appendixB}
We illustrate by means of a simulation the effect of incorrect extinct cohort analysis. To this effect,  $B=20$ years of independent pairs $(x,t)$ were generated with 150 annual exceedances on average, with $x$ uniform on $[0,20]$ and $t$ exponential with mean $\sigma=1/\log 2$.  Three estimates were obtained for each of 1000 replicate datasets: a naive estimate using only extinct cohorts and ignoring the truncation; the maximum likelihood estimate using only extinct cohorts but allowing for truncation; and the estimate using all truncated observations. The left-hand panel shows that the naive estimator has a slight but significant downward bias, unlike the other two, and that the estimator based on all the truncated data is more efficient.  The right-hand panel shows a dataset in which the truncation leads to noticeable non-exponentiality of the lifetimes for the non-extinct cohorts.
\begin{figure}[t!] 
 \centering 
 \includegraphics[width=\textwidth]{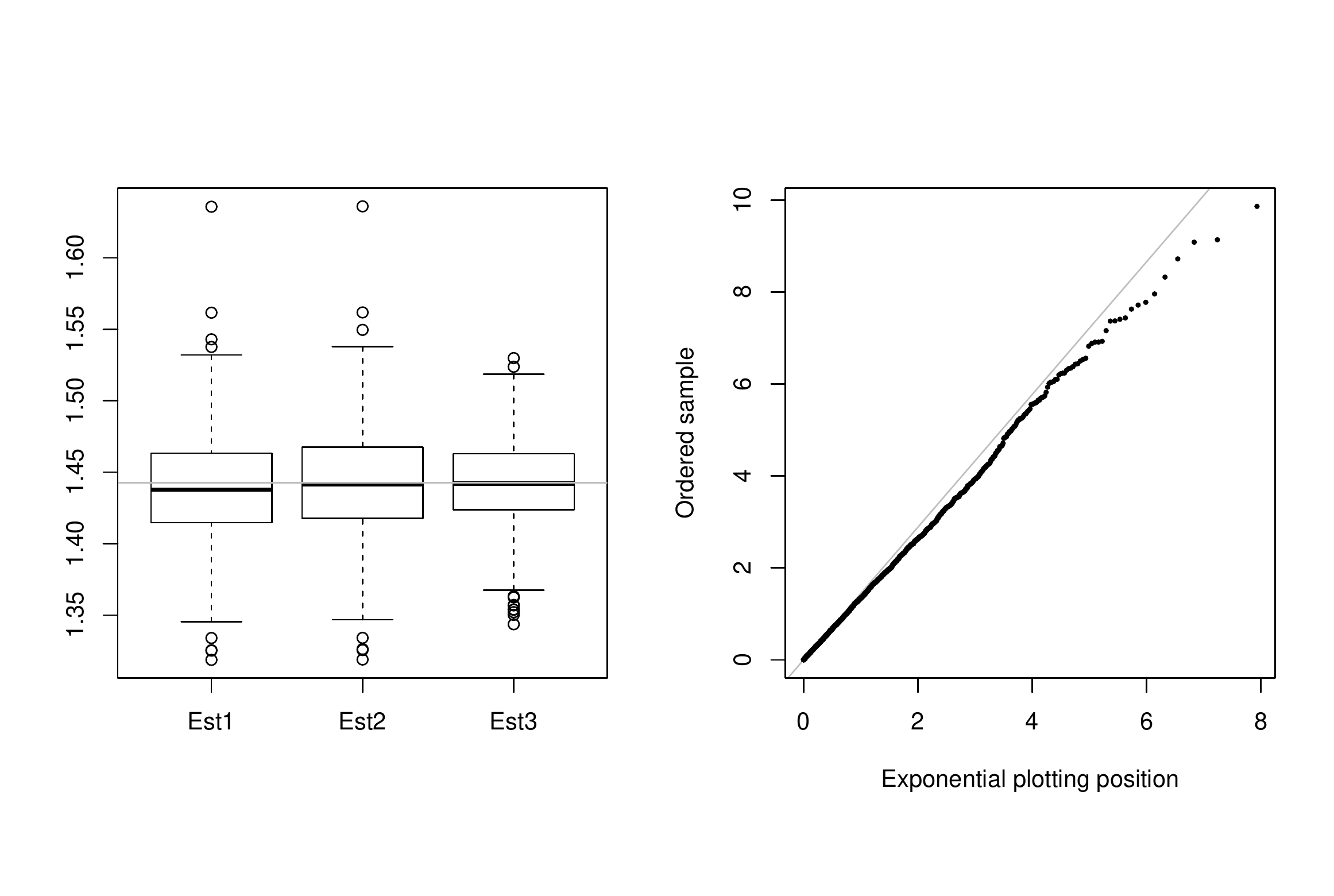}
 \caption{Left: Box-and-whiskers plot of the sampling distribution of three estimators of the exponential hazard for the extinct cohorts (without accounting for truncation, and with truncation) and the full data with truncation. Right: quantile-quantile plot for a selected dataset showing non-exponentiality due to right truncation.}
 \label{ExtinctBias.fig}
 \end{figure}
 
 \section{Extreme value theory: maxima}

The modern formulation of the original large-sample result for the maximum of independent random variables $X_1,\ldots, X_n$ with common distribution function $F$  \citep{Fisher.Tippett:1928} states that if sequences $b_n$ and $a_n>0$  can be chosen so that $T_n=\{\max(X_1,\ldots, X_n)-b_n\}/a_n$ converges to a non-degenerate limiting random variable $T$ as $n\to\infty$, then the distribution of $T$ must be of generalized extreme-value form, i.e.,
\begin{align}\label{gev.eq}
\Pr(T\leq t) =  \begin{cases} 
\exp\left[ - \left\{1 + \xi{(t-\eta)/\tau}\right\}_+^{-1/\xi} \right],&\xi\neq 0,\\ 
\exp\left[ - \exp\left\{-{(t-\eta)/\tau}\right\}\right],&\xi= 0,
\end{cases}
\end{align}
where $a_+=\max(a,0)$ for any real $a$. Here $\eta$ and $\tau$ are respectively location and scale parameters, and the shape parameter $\xi$ determines the form of the upper tail for the limiting distribution and the values attainable by $T$.  If $\xi>0$ then the distribution is said to be of Fr\'echet type, and $T$ can take values in the set $(\eta -\tau/\xi, \infty)$.  If $\xi=0$ then $T$ has a Gumbel distribution, and can take any real value.  If $\xi<0$, then the distribution is said to be of Weibull type, and $T$ takes values in $(-\infty, \eta-\tau/\xi)$.  Thus in this setting the absence of an cap on longevity could be posed as a test of the null hypothesis that $\xi\geq0$, no cap,  against the alternative that $\xi<0$, though it is typically more informative to construct a confidence set for the upper support point $\psi=\eta -\tau/\xi$ for $T$. 

\Cref{gev.eq} can be used to approximate the distribution of the longest lifetime for a population of individuals, and can be used to extrapolate to larger populations.  For example, if~\Cref{gev.eq} corresponds to the maximum of a population of one million persons, then the maximum lifetime for a population of $N$ million persons with the same individual lifetime distributions, $G^N(t)$, is also of the form in~\Cref{gev.eq} with parameters $\eta_N=\eta + \tau(N^\xi-1)/\xi$, $\tau_N=\tau N^\xi$ and $\xi_N=\xi$.  If estimates of the parameters are available, this allows  the distributions of future maxima, or those for other similar populations, to be estimated.  

\section{Penultimate models}
\subsection{Northrop--Coleman piecewise continuous model}
The model proposed by \citet{Northrop.Coleman:2014} consists of truncated generalized Pareto distribution defined over disjoint intervals defined by thresholds $u_1 < \cdots < u_K<u_{K+1}=\infty$ with shape parameters $\xi_1, \ldots, \xi_K$ and scale parameters $\sigma_1,\ldots, \sigma_K$ where $ \sigma_k = \sigma_1 + \sum_{k'=1}^{k-1}\xi_{k'}(u_{k'+1}-u_{k'})$ for $k=2, \ldots, K$. Its  density and distribution functions are
\begin{align*}
 f(t) = \sum_{k=1}^{K} (p_k - p_{k+1}) \dfrac{f_k(t)}{F_k(u_{k+1})}, \qquad F(t) = (1-p_k) + (p_k - p_{k+1})\dfrac{F_k(t)}{F_k(u_{k+1})}, 
\end{align*}
where $f_k$ and $F_k$ denote the density and distribution functions of a generalized Pareto variable  {with location parameter} $u_k$, scale $\sigma_k$ and shape $\xi_k$ and $p_k - p_{k+1} = \Pr(u_k \leq T < u_{k+1})$ is the probability of falling between two adjacent thresholds; $p_k$ can be calculated through the recursion $p_k = \prod_{k'=1}^{k-1} \{1-F_{k'}(u_{{k'}+1})\}$ and samples from $F$ can be obtained by choosing interval $k$ with probability $(p_k - p_{k+1})$ and then simulating a truncated observation in the interval $[u_k, u_{k+1})$.

\subsection{Extended generalized Pareto models}
The hazard function of the extended generalized Pareto model,
\begin{align*}
\sigma^{-1} \dfrac{e^{\beta t/\sigma}}{1 + \xi( e^{\beta t/\sigma} -1)/\beta }, \quad t>0,                                                                                           
\end{align*}
is increasing when $\beta>0$ but has upper asymptote $\beta/(\xi\sigma)$ when $\xi>0$.  When $\xi<0$, the density has support $(0,\sigma\log(1-\beta/\xi)/\beta)$, which reduces to $(0,-\sigma/\xi)$ when $\beta\to0$.   The reciprocal hazard function has derivative 
\begin{align*}
r'(u) = (\xi-\beta)\exp(-\beta u/\sigma), \quad 
\end{align*}
and when $\xi>0$ the penultimate shape parameter depends on the sign of $\xi-\beta$.  When $\xi=0$, the penultimate shape is negative except in the exponential case $\beta=0$. When $\xi<0$, the penultimate shape must be negative (since $\beta\geq 0$), but since $t\to \sigma\log(1-\beta/\xi)/\beta$ implies that $\exp(-\beta t/\sigma)\to (1-\beta/\xi)^{-1}$, we see that $r'(t)\to \xi$.    

A similar example includes the generalized Pareto and Weibull distributions.  The survival function
\begin{align*}
S(t) = \left\{1+\xi \left( t/\sigma\right)^{\beta}\right\}_{+}^{-1/\xi},\quad t>0, \quad \beta,\sigma>0, \xi\in\mathbb{R}, 
\end{align*}
for example, recovers the generalized Pareto model when $\beta=1$ and the Weibull model when $\xi\to 0$. The derivative of the reciprocal hazard, which gives the penultimate shape parameter is 
\begin{align*}
r'(u) = \dfrac{\xi \left(u/\sigma\right)^{\beta} - \beta + 1}{\beta \left(u/\sigma\right)^{\beta}}, \quad u>0, 
\end{align*}
and the shape parameter converges to $\xi/\beta$  as $u \to \infty$. The parameters of this model are unlikely to be numerically identifiable, but one could test the adequacy of the Weibull and generalized Pareto distributions through score tests for the hypotheses $ \xi=0$ and $\beta=1$.

\section{Quantile-quantile plots for truncated data} \label{sec:appendixD}

A quantile-quantile plot compares the empirical quantiles of a random sample with their expected values under a distribution $F_0$, often fitted to the sample.  Let $y_1,\ldots, y_n$ denote the ordered sample values and let $v_1,\ldots, v_n$ denote the corresponding empirical positions; typically $v_j = F_0^{-1}\{F_n(y_j)\}$, where $F_n$ is a nonparametric estimate of the distribution; in the absence of censoring we generally take $F_n(y_j) = \mathrm{rank}(y_j)/(n+1)$.

With right-censored data, $F_n$ can be based on the product-limit estimator, but truncated data from the distribution $F_0$ need a different treatment, because the cumulative distribution distribution for each observation depends on its truncation interval. If the observations are truncated in a single interval, this gives
\begin{align}\label{qq.eq}
 {F}^{(i)}_0(x) = \dfrac{F_0(x) - F_0(a_i)}{F_0(b_i)-F_0(a_i)}, \quad a_i<x<b_i;
\end{align}
we define analogously ${F}^{(i)}_n(x)$ as the empirical distribution, accounting for truncation limits of the $i$th record. 
There are two strategies for displaying the data on the scale of $F_0$.  The first is to map the $y_j$ to uniform positions via~\eqref{qq.eq} and to display transformed empirical quantiles ${y}_i = F^{-1}_0 \{ F_0^{(i)}(y_i)\}$ on the $y$-axis. These could be plotted against ordinary plotting positions $v_i =  F_0^{-1}\{F_n({y}_i)\}$ or against $u_i = F_0^{-1}\{{F}_n^{(i)}(y_i)\}$; while use of $u_i$ is formally correct, use of $v_i$ has the advantage that the ranks of the $v_i$ and the ${y}_i$ are the same, which is not guaranteed if one uses $u_i$, though plotting the pairs $(u_i, {y}_i)$ should still give a straight line.
The second strategy is to leave the $y_i$ unchanged, but to adapt the plotting positions to 
\begin{align}
 v_i= {F}_0^{(i)}{}^{-1}\{{F}_n^{(i)}(y_i)\}=F_0^{-1}\left[F_0(a_i) +\{F_0(b_i)-F_0(a_i)\}\frac{F_n(y_i) - F_n(a_i)}{F_n(b_i)-F_n(a_i)}\right] ;\label{eq:pos}
\end{align}
these need not be ordered.

Quantile-quantile plots can be hard to interpret without some indication of their variability.  This can be obtained by sampling observations from the null model $F_0$ with the same truncation bounds (including cases with right-censoring, which are used to estimate the model despite not being displayed on the graphic). However, when we re-estimate $F_n$ and $F_0$ for each bootstrap sample, both $x$ and $y$ plotting positions change.
% , so it is not obvious how to obtain approximate confidence intervals. Two possibilities are to simulate new data from $F_0$ keeping the bounds $(a_i, b_i)$ for truncation and censoring, obtaining an estimate $\widehat{F}_0^{(b)}$ for each simulated dataset, or simulating new parameter values by normal approximation to the maximum likelihood estimators of the parameters of $F_0$.   In general calculating $\widehat{F}_n^{(b)}$ requires an EM algorithm and could be costly in large samples. 
%   \begin{itemize}
%   \item For the quantile-quantile plot, we could keep the same plotting positions $y_i$ on the $y$-axis and only recalculate the plotting positions on the $x$-axis associated to $y_i$ in eq.~\ref{eq:pos} where $F_0$ and $F_n$ would be substituted with $\widehat{F}_0^{(b)}$ and ${F}_n^{(n)}$.
%   \item For a probability-probability plot, both $x$ and $y$ plotting positions would change unless we do not recompute $F_n$ with each bootstrap sample.
% \end{itemize}
In Figure~6, we ignored the dependence between ordered observations of the same bootstrap sample and pooled all of the data together. The intervals were estimated using quantile regression with level $\tau=0.05$ and $\tau=0.95$ through constrained regression splines to obtain the confidence intervals (with the $\mathsf{R}$ package \texttt{cobs}). The overall empirical coverage of this 90\% pointwise confidence interval is 81\%.

\section{Analysis of the \IDL{} 2021 database} \label{sec:appendixF}

Are there differences in survival for supercentenarians of different countries or gender? \Cref{table:groups} provides a breakdown of people aged above 110 in the \IDL{} database \citep{IDL:2021} by gender and region. One way of testing for such differences is to fit a parametric model for each gender/region separately and compare the fit with the same model fitted to the pooled data, corresponding to the null hypothesis of equal distribution. The asymptotic null distribution of the likelihood ratio statistic is $\chi^2_{(m-1)k}$, where $m$ is the number of groups under comparison and $k$ is the number of parameters of the model. We computed the $p$-values for comparisons between women and men and and between regions (\Cref{table:women-men,table:countries}) but found none. If there was no difference, we would expect the $p$-values to be uniform: with independent datasets, this would probably imply larger variability than is visible in \Cref{table:women-men,table:countries} but these data overlap due to our definition. These calculations were performed  using the \textsf{MATLAB} toolbox using the \IDL{} data currently available online.

Table~\ref{table:early-late} shows that for the North America data survival is larger for individuals who have died later, with a accompanying small $p$-value. However differences go the other way for the Europe data, and there is little evidence of a time trend in the combined data set, World.  

We can also check the evidence for an upper bound. \Cref{table:test-exp} gives no evidence of deviation from the exponential model.  Accordingly, the confidence intervals all include both negative and positive values of $\xi$, so a finite endpoint is not incompatible with the data, but it would lead to a very large upper bound, as shown in Section~6 of the paper.

We can also consider countries with semisupercentenarians other than those analyzed in the paper. We selected all individuals who died in [$\max(c_1, d_1), \min(c_2, d_2)$] to avoid dealing with the double interval-truncation. According to \Cref{table:thresholds}, there is no evidence against a constant hazard from age 106 onwards for the group of countries and subnational units consisting of Austria, Belgium, Denmark, Germany and Quebec.

\begin{table}[htbp!]
\centering
\caption{Datasets used in the analysis, with sample size broken down per gender.}
% For the \IDL{}, persons with validation level B are excluded.}
\label{table:groups}
\vspace*{3mm}
\begin{tabular}{l p{7.5cm} c c}
\toprule
ID & list of countries &  women & men
\\ 
\midrule
	North Europe & Austria, Belgium, Denmark, Finland, Germany, Norway, Sweden, England and Wales    
														  & 208 & 16 \\
	South Europe & France, Spain                          & 282 & 17 \\
	Europe & North and South Europe                       & 490 & 33 \\
	North America & USA, Quebec                           & 468 & 47 \\ % (Canada)
	World & Europe\& North America                         & 958 & 80 \\
\bottomrule
\end{tabular}
\end{table}

\begin{table}[htbp!]
\centering
\caption{$p$-values for likelihood ratio tests comparing different generalized Pareto models (middle) and different exponential models (right) for women and men against the null hypothesis of the same distribution for women and men, for \IDL{} supercentenarian from different regions.}
\label{table:women-men}
\vspace*{3mm}
\begin{tabular}{lcc}
\toprule
ID  &  generalized Pareto & exponential
\\
\midrule
	North Europe    & 0.74 & 0.6\phantom{2} \\
	South Europe    & 0.67 & 0.52 \\
	Europe          & 0.98 & 0.9\phantom{2} \\
	North America   & 0.83 & 0.70 \\
	World           & 0.88 & 0.78 \\
\bottomrule
\end{tabular}
\end{table}

\begin{table}[htbp!]
\centering
\caption{$p$-values for Wald tests comparing different exponential distributions (last column) for the  first half ($\sigma_1)$ and second half ($\sigma_2)$ of the data, with sample sizes and years of death for the observation scheme.}
\vspace*{3mm}
\begin{tabular}{l c c c c c ccl}
\toprule
ID & death date & $n_1$ & $\widehat{\sigma}_1$ & death date & $n_2$ & $\widehat{\sigma}_2$ & \multicolumn{1}{c}{$p$-value}
\\
\midrule
	North Europe                  & 1968--2007 & 116 & 1.45 & 2007--2018 & 108 & 1.16 & 0.17\phantom{2}  \\
	South Europe                  & 1973--2010 & 149 & 1.50 & 2010--2018 & 150 & 1.47 & 0.88\phantom{2}  \\
	Europe                        & 1968--2009 & 260 & 1.34 & 2009--2018 & 244 & 1.48 & 0.38\phantom{2}  \\
	North America                 & 1962--1998 & 253 & 1.15 & 1998--2011 & 262 & 1.64 & 0.002 \\
	World                         & 1962--2002 & 507 & 1.37 & 2002--2018 & 531 & 1.40  & 0.71\phantom{2}  \\
\bottomrule
\end{tabular}
\label{table:early-late}
\end{table}

\begin{table}[htbp!]
\centering
\caption{$p$-values for the likelihood ratio tests comparing different generalized Pareto  (middle) and exponential  (right) distributions between groups of countries for \IDL{} supercentenarian against the null of common distribution for both groups of countries.}
\label{table:countries}
\vspace*{3mm}
\begin{tabular}{l c c}
\toprule
ID & generalized Pareto & exponential
\\
\midrule 
	North Europe vs South Europe & 0.6\phantom{2}  & 0.7\phantom{2}  \\
	Europe vs North America      & 0.57 & 0.66 \\
\bottomrule
\end{tabular}
\end{table}

\begin{table}[htbp!]
\centering
\caption{Estimates ($95$\% Wald-based confidence intervals) of the shape parameter of the generalized Pareto distribution, with $p$-values for likelihood ratio test of the null hypothesis of exponential distribution ($\xi=0$).}
\label{table:test-exp}
\vspace*{3mm}
\begin{tabular}{l l c }
\toprule
ID & shape & $p$-value\\
\midrule
		North Europe    & $-$0.03 \; \; ($-$0.21, 0.16) & 0.58 \\ 
		South Europe    &\phantom{$-$}0.05 \; \; ($-$0.09, 0.18) & 0.46 \\  
		Europe          &\phantom{$-$}0.01 \; \; ($-$0.08, 0.11) & 0.77 \\ 
		North America   & $-$0.04 \; \; ($-$0.13, 0.05) & 0.39 \\ 
		World           & $-$0.01 \; \; ($-$0.07, 0.05) & 0.76\\
\bottomrule
\end{tabular}
\end{table}
 \begin{table}[t!]
\centering
\caption{ Estimates (standard errors) of the scale parameter $\sigma$ of fitted exponential distributions and $p$-value for likelihood ratio tests of the exponential model as a submodel of the generalized Pareto model, for the \IDL{} 2021 semi- and supercentenarian datasets for Austria, Belgium, Denmark, Germany and Quebec, as a function of threshold, with number of threshold exceedances ($n_u$)}.

\vspace*{3mm}
\begin{tabular}{l c c c c c ccc}
\toprule
threshold   & 105         & 106         & 107          & 108         & 109          & 110
\\
\midrule
$\sigma$ & 1.53 (0.04) & 1.43 (0.05) & 1.35 (0.07)  & 1.26 (0.09) & 1.27 (0.15)  & 1.10 (0.19)   \\
$p$-value    & 0.02        & 0.19        & 0.29         & 0.54        & 0.49         & 0.95           \\
$n_u$                        & 2\ 156       & 1\ 104       & 537          & 244         & 102           & 45             \\
\bottomrule
\end{tabular}
\label{table:thresholds}
 \end{table}
\end{document}